# Capacity of Steganographic Channels

Jeremiah J. Harmsen, *Member, IEEE,* William A. Pearlman, *Fellow, IEEE,*

*Abstract*—This work investigates a central problem in steganography, that is: How much data can safely be hidden without being detected? To answer this question, a formal definition of steganographic capacity is presented. Once this has been defined, a general formula for the capacity is developed. The formula is applicable to a very broad spectrum of channels due to the use of an information-spectrum approach. This approach allows for the analysis of arbitrary steganalyzers as well as non-stationary, non-ergodic encoder and attack channels.

After the general formula is presented, various simplifications are applied to gain insight into example hiding and detection methodologies. Finally, the context and applications of the work are summarized in a general discussion.

*Index Terms*—Steganographic capacity, stego-channel, steganalysis, steganography, information theory, information spectrum

## I. Introduction

### A. Background

SHANNON'S pioneering work provides bounds on the amount of information that can be transmitted over a noisy channel. His results show that capacity is an intrinsic property of the channel itself. This work takes a similar viewpoint in seeking to find the amount of information that may be transferred over a stego-channel as seen in Figure 1.

The stego-channel is equivalent to the classic channel with the addition of the detection function and attack channel. For the classic channel, a transmission is considered successful if the decoder properly determines which message the encoder has sent. In the stego-channel, a transmission is successful only if the decoder properly determines the sent message and the detection function is not triggered.

This additional constraint on the channel use leads to the fundamental view that *the capacity of a stego-channel is an intrinsic property of both the channel and the detection function*. That is to say, the properties of the detection function influence the capacity just as much as the noise in the channel.

### B. Previous Work

There have been a number of applications of information theory to the steganographic capacity problem[1], [2], [3]. These works give capacity results under distortion constraints on the hider as well as active adversary. The additional constraint that the stego-signal retains the same distribution as the cover-signal serves as the steganalysis detection function.

Somewhat less work exists exploring capacity with arbitrary detection functions. These works are written from a steganalysis perspective[4], [5] and accordingly give heavy consideration to the detection function.

This work differs from previous work in a number of aspects. Most notable is the use of information-spectrum methods that allow for the analysis of arbitrary detection algorithms and channels. This eliminates the need to restrict interest to detection algorithms that operate on sample averages or behave consistently. Instead, the detection functions may be instantaneous, meaning the properties of a detector for $n$ samples need not have any relation to the same detector for $n+1$ samples. Additionally, the typical restriction that the channel under consideration be consistent, ergodic or stationary is also lifted.

Another substantial difference is the presence of noise *before* the detector. This placement enables the modeling of common signal processing distortions such as compression, quantization, etc. The location of the noise adds complexity not only because of confusion at the decoder, but also because a signal, carefully crafted to avoid detection, may be corrupted into one that will trigger the detector.

Finally, the consideration of a cover-signal and distortion constraint in the encoding function is omitted. This is due to the view that steganographic capacity is a property of the channel and the detection function. This viewpoint, along with the above differences, make a direct comparison to previous work somewhat difficult, although possible with a number of simplifications explored in Section V.

### C. Groundwork

This chapter lays the groundwork for determining the amount of information that may be transferred over the channel shown in Figure 1. Here, the adversary's goal is to disrupt any steganographic communication between the encoder and decoder. To accomplish this a steganalyzer is used to detect steganographic messages and an attack function is used to corrupt undetected messages.

We now formally define each of the components in the system, beginning with the random variable notation.

*1) Random Variables:* Random variables are denoted by capital letters, e.g. $X$. Realizations of these random variables are denoted as lowercase letters, e.g. $x$. Each random variable is defined over a domain denoted with a script $\mathcal{X}$. A sequence of $n$ random variables is denoted with $X^n = (X_1, \ldots, X_n)$. Similarly, an $n$-length sequence of random variable realizations is denoted $\mathbf{x} = (x_1, \ldots, x_n) \in \mathcal{X}^n$. The probability of $X$ taking value $x \in \mathcal{X}$ is $p_X(x)$.

Following a signal through Figure 1, we begin in the space of $n$-length stego-signals denoted $\mathcal{X}^n$. The signal then

This work was carried out at Rensselaer Polytechnic Institute and was supported by the Air Force Research Laboratory, Rome, NY.

J. Harmsen is now with Google Inc. in Mountain View, CA 94043, USA; E-mail: jeremiah@google.com.

W. Pearlman is with the Elec. Comp. and Syst. Engineering Dept., Rensselaer Polytechnic Institute, Troy, NY 12180-3590, USA; E-mail: pearlw@ecse.rpi.edu.

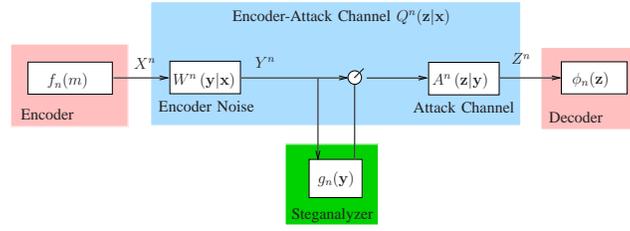

Fig. 1. Steganographic Channel

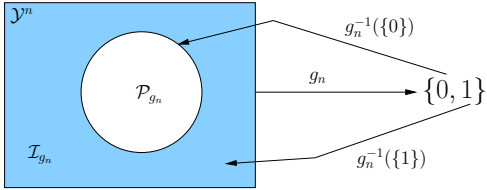

Fig. 2. Permissible and Impermissible Sets

undergoes some distortion as it travels through the encoder-channel. This results in an element from the corrupted stego-signal space of $\mathcal{Y}^n$. Finally, the signal is attacked to produce the attacked stego-signal in space $\mathcal{Z}^n$.

*2) Steganalyzer:* The *steganalyzer* is a function $g_n : \mathcal{Y}^n \to \{0,1\}$ that classifies a sequence of signals from $\mathcal{Y}^n$ into one of two categories: containing steganographic information and not containing steganographic information. The function is defined as follows for all $\mathbf{y} \in \mathcal{Y}^n$,

$$g_n(\mathbf{y}) = \begin{cases} 1, & \text{if } \mathbf{y} \text{ is steganographic} \\ 0, & \text{if } \mathbf{y} \text{ is not steganographic} \end{cases} \quad (1)$$

The specific type of function may be that of support vector machine or a Bayesian classifier, etc.

A *steganalyzer sequence* is denoted as,

$$\mathbf{g} := \{g_1, g_2, g_3, \ldots\}, \quad (2)$$

where $g_n : \mathcal{Y}^n \to \{0, 1\}$.

The set of all $n$ length steganalyzers is denoted $\mathcal{G}_n$.

*3) Permissible Set:* For any steganalyzer $g_n$, the space of signals $\mathcal{Y}^n$ is split into the permissible set and the impermissible set.

The *permissible set* $\mathcal{P}_{g_n} \subseteq \mathcal{Y}^n$ is the inverse image of 0 under $g_n$,

$$\mathcal{P}_{g_n} := g_n^{-1}(\{0\}) = \{\mathbf{y} \in \mathcal{Y}^n : g_n(\mathbf{y}) = 0\}. \quad (3)$$

The permissible set is the set of all signals of $\mathcal{Y}^n$ that the given steganalyzer, $g_n$ will classify as non-steganographic.

Since each steganalyzer has a binary range, a steganalyzer sequence may be completely described by a sequence of permissible sets. To denote a steganalyzer sequence in such a way the following notation is used,

$$\mathbf{g} \cong \{\mathcal{P}_1, \mathcal{P}_2, \mathcal{P}_3, \ldots\},$$

where $\mathcal{P}_n \subseteq \mathcal{Y}^n$ is the permissible set for $g_n$.

*4) Impermissible Set:* The *impermissible set* $\mathcal{I}_{g_n} \subseteq \mathcal{Y}^n$ is the inverse image of 1 under $g_n$,

$$\mathcal{I}_{g_n} := g_n^{-1}(\{1\}) = \{\mathbf{y} \in \mathcal{Y}^n : g_n(\mathbf{y}) = 1\}. \quad (4)$$

For a given $g_n$ the impermissible set is the set of all signals in $\mathcal{Y}^n$ that $g_n$ will classify as steganographic.

*Example 1:* Consider the illustrative sum steganalyzer defined for the binary channel outputs ($\mathcal{Y} = \{0,1\}$). The steganalyzer is defined for $\mathbf{y} = (y_1, \ldots, y_n)$ as,

$$g_n(\mathbf{y}) = \begin{cases} 1, & \text{if } \sum_{i=1}^{n} y_i > \lfloor \frac{n}{2} \rfloor \\ 0, & \text{else} \end{cases} \quad (5)$$

The permissible sets for $n = 1, 2, 3, 4$ are shown in Table I.

TABLE I
SUM STEGANALYZER PERMISSIBLE SETS

| | |
|---|---|
| $\mathcal{P}_1 =$ | $\{(0)\}$ |
| $\mathcal{P}_2 =$ | $\{(0,0),(0,1),(1,0)\}$ |
| $\mathcal{P}_3 =$ | $\{(0,0,0),(1,0,0),(0,1,0),(0,0,1)\}$ |
| $\mathcal{P}_4 =$ | $\{(0,0,0,0),(1,0,0,0),(0,1,0,0),(0,0,1,0),(0,0,0,1),$ $(1,1,0,0),(1,0,1,0),(1,0,0,1),(0,1,1,0),(0,1,0,1),(0,0,1,1)\}$ |

*5) Memoryless Steganalyzers:* A *memoryless steganalyzer*, $\mathbf{g} = \{g_n\}_{n=1}^{\infty}$ is one where each $g_n$ is defined for $\mathbf{y} = (y_1, y_2, \ldots, y_n)$ as,

$$g_n(\mathbf{y}) = \begin{cases} 1, & \text{if } \exists i \in \{1, 2, \ldots, n\} \text{ such that } g(y_i) = 1 \\ 0, & \text{if } g(y_i) = 0 \ \forall \ i \in \{1, 2, \ldots, n\} \end{cases} \quad (6)$$

where $g \in \mathcal{G}_1$ is said to specify $g_n$ (and $\mathbf{g}$). To denote a steganalyzer sequence is memoryless the following notation will be used $\mathbf{g} = \{g\}$.

The analysis of the memoryless steganalyzer is motivated by the current real world implementation of detection systems. As an example we may consider each $y_i$ to be a digital image sent via email. When sending $n$ emails, the hider attaches one of the $y_i$'s to each message. The entire sequence of images is considered to be $\mathbf{y}$. Typically steganalyzers do not make use of entire sequence $\mathbf{y}$. Instead, each image is sequentially processed by a given steganalyzer $g$, where if any of the $y_i$ trigger the detector the entire sequence of emails is treated as steganographic.

For a memoryless steganalyzer $g_n$ defined by $g$, the permissible set of $g_n$ is defined by the $n$-dimensional product of $\mathcal{P}_g$,

$$\mathcal{P}_{g_n} = \underbrace{\mathcal{P}_g \times \mathcal{P}_g \times \cdots \times \mathcal{P}_g}_{n}. \quad (7)$$

## D. Channels

We now define two channels. The first models inherent distortions occurring between the encoder and detection function, such as the compression of the stego-signal. The second models a malicious attack by an active adversary such as a cropping or additive noise. Both of these distortions are considered to be outside the control of the encoder.

*1) Encoder-Noise Channel:* The *encoder-noise channel* is denoted as $W^n$ where $W^n : \mathcal{Y}^n \times \mathcal{X}^n \to [0,1]$ and has the following property for all $\mathbf{x} \in \mathcal{X}^n$,

$$W^n(\mathbf{y}|\mathbf{x}) := \Pr\{Y^n = \mathbf{y} | X^n = \mathbf{x}\}.$$

The channel represents the conditional probabilities of the steganalyzer receiving $\mathbf{y} \in \mathcal{Y}^n$ when $\mathbf{x} \in \mathcal{X}^n$ is sent.

The random variable, $Y$ resulting from transmitting $X$ through the channel $W$ will be denoted as $X \xrightarrow{W} Y$.

We denote an *arbitrary encoder-noise channel* as the sequence of transition probabilities,

$$\mathbf{W} := \{W^1, W^2, W^3, \ldots\}.$$

*2) Attack Channel:* The attack function maps $A^n : \mathcal{Y}^n \to \mathcal{Z}^n$ as,

$$A^n(\mathbf{z}|\mathbf{y}) = \Pr\{Z^n = \mathbf{z} | Y^n = \mathbf{y}\}. \tag{8}$$

The attack channel may be deterministic or probabilistic.

Similar to the encoder-noise channel, we denote an *arbitrary attack channel* as the sequence of transition probabilities,

$$\mathbf{A} := \{A^1, A^2, A^3, \ldots\}.$$

*3) Encoder-Attack Channel:* The *encoder-attack channel* or *channel* is a function $Q^n : \mathcal{X}^n \to \mathcal{Z}^n$, defined to model the effect of both the encoder-noise and attack channel,

$$Q^n(\mathbf{z}|\mathbf{x}) = \sum_{\mathbf{y} \in \mathcal{Y}^n} A^n(\mathbf{z}|\mathbf{y}) W^n(\mathbf{y}|\mathbf{x}). \tag{9}$$

The specification of $Q^n$ by $A^n$ and $W^n$ is denoted $Q^n = A^n \circ W^n$.

The *arbitrary encoder-attack channel* is a sequence of transition probabilities,

$$\mathbf{Q} = \{Q^1, Q^2, Q^3, \ldots\}. \tag{10}$$

We will express the relation between the encoder-noise channel, attack channel and encoder-attack channel as $\mathbf{Q} = \mathbf{A} \circ \mathbf{W}$.

*4) Memoryless Channels:* In the case where channel distortions act independently and identically on each input letter $x_i$, we say it is a *memoryless channel*. In this instance the $n$-length transition probabilities can be written as,

$$W^n(\mathbf{y}|\mathbf{x}) = \prod_{i=1}^{n} W(y_i|x_i), \tag{11}$$

where $W$ is said to define the channel. To denote a channel is memoryless and defined by $W$ we will write $\mathbf{W} = \{W\}$.

## E. Encoder and Decoder

The purpose of the encoder and decoder is to transmit and receive information across a channel. The information to be transferred is assumed to be from a uniformly distributed message set denoted $\mathcal{M}_n$, with a cardinality of $M_n$.

The *encoding function* maps a message to a stego-signal, i.e. $f_n : \mathcal{M}_n \to \mathcal{X}^n$. The element of $\mathcal{X}^n$ to which the $i$th message maps is called the *codeword* for $i$ and is denoted, $\mathbf{u}_i$. The collection of codewords, $\mathcal{C}_n = \{\mathbf{u}_1, \ldots, \mathbf{u}_{M_n}\}$ is called the *code*. The *rate*, $R_n$ of an encoding function is given as $\frac{1}{n} \log M_n$.

The *decoding function*, $\phi_n : \mathcal{Z}^n \to \mathcal{M}_n$, maps a corrupted stego-signal to a message. The decoder is defined by the set of *decoding regions* for the each message. The decoding regions, $\mathcal{D}_1, \ldots, \mathcal{D}_{M_n}$, are disjoint sets that cover $\mathcal{Z}^n$ and defined such that,

$$\phi_n^{-1}(\{m\}) = \mathcal{D}_m$$
$$:= \{F \subseteq \mathcal{Z}^n : \phi_n(\mathbf{z}) = m, \ \forall \ \mathbf{z} \in F\},$$

for $m = 1, \ldots, M_n$.

Next, two important terms are presented that allow for the analysis of steganographic systems. The first is the probability the decoder makes a mistake, called the probability of error. The second is the probability the steganalyzer is triggered, called the probability of detection. In both cases they are calculated for a given code $\mathcal{C} = \{\mathbf{u}_1, \ldots, \mathbf{u}_{M_n}\}$, encoder-channel $W^n$, attack-channel $A^n$ and impermissible set $\mathcal{I}_{g_n}$ (corresponding to some $g_n$).

The *probability of error* in decoding the message can be found as,

$$\epsilon_n = \frac{1}{M_n} \sum_{i=1}^{M_n} Q^n(\mathcal{D}_i^c | \mathbf{u}_i), \tag{12}$$

where $Q^n = A^n \circ W^n$.

Similarly the *probability of detection* for the steganalyzer is calculated as,

$$\delta_n = \frac{1}{M_n} \sum_{i=1}^{M_n} W^n(\mathcal{I}_{g_n} | \mathbf{u}_i). \tag{13}$$

## F. Stego-Channel

A *steganographic channel* or *stego-channel* is a triple $(\mathbf{W}, \mathbf{g}, \mathbf{A})$, where $\mathbf{W}$ is an arbitrary encoder-noise channel, $\mathbf{g}$ is a steganalyzer sequence, and $\mathbf{A}$ is an arbitrary attack channel. To reinforce the notion that a stego-channel is defined by a sequence of triples we will typically write $(\mathbf{W}, \mathbf{g}, \mathbf{A}) = \{(W^n, g_n, A^n)\}_{n=1}^{\infty}$.

*1) Discrete Stego-Channel:* A *discrete stego-channel* is one where at least one of the following holds:

$$|\mathcal{X}| < \infty, \quad |\mathcal{Y}| < \infty, \quad |\mathcal{Z}| < \infty, \quad \text{or } |\mathcal{P}_{g_n}| < \infty \ \forall n.$$

*2) Discrete Memoryless Stego-Channel:* A *discrete memoryless stego-channel* (DMSC) is a stego-channel where,

1) $(\mathbf{W}, \mathbf{g}, \mathbf{A})$ is discrete
2) $\mathbf{W}$ is memoryless
3) $\mathbf{g}$ is memoryless
4) $\mathbf{A}$ is memoryless

A DMSC is said to be defined by the triple $(W, g, A)$ and will be denoted $(\mathbf{W}, \mathbf{g}, \mathbf{A}) = \{(W, g, A)\}$.

### G. Steganographic Capacity

The secure capacity tells us how much information can be transferred with arbitrarily low probabilities of error and detection.

An $(n, M_n, \epsilon_n, \delta_n)$-code (for a given stego-channel) consists of an encoder and decoder. The encoder and decoder are capable of transferring one of $M_n$ messages in $n$ uses of the channel with an average probability of error of less than (or equal to) $\epsilon_n$ and a probability of detection of less than (or equal to) $\delta_n$.

*1) Secure Capacity:* A rate $R$ is said to be *securely achievable* for a stego-channel $(\mathbf{W}, \mathbf{g}, \mathbf{A}) = \{(W^n, g_n, A^n)\}_{n=1}^{\infty}$, if there exists a sequence of $(n, M_n, \epsilon_n, \delta_n)$-codes such that:

1) $\lim_{n\to\infty} \epsilon_n = 0$
2) $\lim_{n\to\infty} \delta_n = 0$
3) $\liminf_{n\to\infty} \frac{1}{n} \log M_n \geq R$

The *secure capacity* of a stego-channel $(\mathbf{W}, \mathbf{g}, \mathbf{A})$ is denoted as $C(\mathbf{W}, \mathbf{g}, \mathbf{A})$. This is defined as the supremum of all securely achievable rates for $(\mathbf{W}, \mathbf{g}, \mathbf{A})$.

### H. $(\epsilon, \delta)$-Secure Capacity

A rate $R$ is said to be $(\epsilon, \delta)$-*securely achievable* for a stego-channel $(\mathbf{W}, \mathbf{g}, \mathbf{A}) = \{(W^n, g_n, A^n)\}_{n=1}^{\infty}$, if there exists a sequence of $(n, M_n, \epsilon_n, \delta_n)$-codes such that:

1) $\limsup_{n\to\infty} \epsilon_n \leq \epsilon$
2) $\limsup_{n\to\infty} \delta_n \leq \delta$
3) $\liminf_{n\to\infty} \frac{1}{n} \log M_n \geq R$

## II. SECURE CAPACITY FORMULA

### A. Information-Spectrum Methods

The information-spectrum method[6], [7], [8], [9], [10] is a generalization of information theory created to apply to systems where either the channel or its inputs are not necessarily ergodic or stationary. Its use is required in this work because the steganalyzer is not assumed to have any ergodic or stationary properties.

The information-spectrum method uses the *general source* (also called *general sequence*) defined as,

$$\mathbf{X} := \left\{ X^n = (X_1^{(n)}, X_2^{(n)}, \ldots, X_n^{(n)}) \right\}_{n=1}^{\infty}, \quad (14)$$

where each $X_m^{(n)}$ is a random variable defined over alphabet $\mathcal{X}$. It is important to note that the general source makes no assumptions about consistency, ergodicity, or stationarity.

The information-spectrum method also uses two novel quantities defined for sequences of random variables, called the lim sup and lim inf in probability.

The *limsup in probability* of a sequence of random variables, $\{Z_n\}_{n=1}^{\infty}$ is defined as,

$$\text{p-}\limsup Z_n := \inf \left\{ \alpha : \lim_{n\to\infty} \Pr\{Z_n > \alpha\} = 0 \right\}.$$

Similarly, the *liminf in probability* of a sequence of random variables, $\{Z_n\}_{n=1}^{\infty}$ is,

$$\text{p-}\liminf Z_n := \sup \left\{ \beta : \lim_{n\to\infty} \Pr\{Z_n < \beta\} = 0 \right\}.$$

The *spectral sup-entropy rate* of a general source $\mathbf{X} = \{X^n\}_{n=1}^{\infty}$ is defined as,

$$\overline{H}(\mathbf{X}) := \text{p-}\limsup_{n\to\infty} \frac{1}{n} \log \frac{1}{p_{X^n}(X^n)}. \quad (15)$$

Analogously, the *spectral inf-entropy rate* of a general source $\mathbf{X} = \{X^n\}_{n=1}^{\infty}$ is defined as,

$$\underline{H}(\mathbf{X}) := \text{p-}\liminf_{n\to\infty} \frac{1}{n} \log \frac{1}{p_{X^n}(X^n)}. \quad (16)$$

The spectral entropy rate has a number of natural properties such as for any $\mathbf{X}$, $\overline{H}(\mathbf{X}) \geq \underline{H}(\mathbf{X}) \geq 0$ [6, Thm. 1.7.2].

The *spectral sup-mutual information rate* for the pair of general sequences $(\mathbf{X}, \mathbf{Y}) = \{(X^n, Y^n)\}_{n=1}^{\infty}$ is defined as,

$$\overline{I}(\mathbf{X}; \mathbf{Y}) := \text{p-}\limsup_{n\to\infty} \frac{1}{n} i(X^n; Y^n), \quad (17)$$

where,

$$i(X^n; Y^n) := \log \frac{p_{Y^n|X^n}(Y^n|X^n)}{p_{Y^n}(Y^n)}. \quad (18)$$

Likewise the *spectral inf-mutual information rate* for the pair of general sequences $(\mathbf{X}, \mathbf{Y}) = \{(X^n, Y^n)\}_{n=1}^{\infty}$ is defined as,

$$\underline{I}(\mathbf{X}; \mathbf{Y}) := \text{p-}\liminf_{n\to\infty} \frac{1}{n} i(X^n; Y^n). \quad (19)$$

### B. Information-Spectrum Results

This section lists some of the fundamental results from information-spectrum theory [6] that will be used in the remainder of the paper.

$$\underline{H}(\mathbf{X}) \leq \liminf_{n\to\infty} \frac{1}{n} H(X^n) \quad (20)$$

$$\underline{I}(\mathbf{X}; \mathbf{Y}) \leq \overline{H}(\mathbf{Y}) - \overline{H}(\mathbf{Y}|\mathbf{X}) \quad (21)$$

$$\underline{I}(\mathbf{X}; \mathbf{Y}) \geq \underline{H}(\mathbf{X}) - \overline{H}(\mathbf{Y}|\mathbf{X}) \quad (22)$$

### C. Secure Sequences

*1) Secure Input Sequences:* For a given stego-channel $(\mathbf{W}, \mathbf{g}, \mathbf{A})$, a general source $\mathbf{X} = \{X^n\}_{n=1}^{\infty}$ is called $\delta$-secure if the resulting $\mathbf{Y} = \{Y^n\}_{n=1}^{\infty}$ satisfies,

$$\limsup_{n\to\infty} \Pr\{g_n(Y^n) = 1\} \leq \delta, \quad (23)$$

or either of the following equivalent conditions,

$$\limsup_{n\to\infty} p_{Y^n}(\mathcal{I}_{g_n}) \leq \delta, \quad (24)$$

or

$$\liminf_{n\to\infty} p_{Y^n}(\mathcal{P}_{g_n}) \geq 1 - \delta. \quad (25)$$

The set, $\mathcal{S}_\delta$, of all general sources that are $\delta$-secure is defined as,

$$\mathcal{S}_\delta := \left\{ \mathbf{X} : \limsup_{n\to\infty} \sum_{\mathbf{x}\in\mathcal{X}^n} W^n(\mathcal{I}_{g_n}|\mathbf{x}) p_{X^n}(\mathbf{x}) \leq \delta \right\}, \quad (26)$$

where $\mathbf{X} = \{X^n\}_{n=1}^{\infty}$.

The set for $\delta = 0$ is called *secure input set* and denoted $\mathcal{S}_0$.

*2) Secure Output Sequences:* For a given steganalyzer sequence $\mathbf{g} = \{g_n\}_{n=1}^{\infty}$, a general sequence $\mathbf{Y} = \{Y^n\}_{n=1}^{\infty}$ is called $\delta$-secure if,

$$\limsup_{n \to \infty} \Pr\{g_n(Y^n) = 1\} \leq \delta, \quad (27)$$

The set, $\mathcal{T}_\delta$, of all $\delta$-secure general output sequences is defined as,

$$\mathcal{T}_\delta := \left\{\mathbf{Y} = \{Y^n\}_{n=1}^{\infty} : \limsup_{n \to \infty} p_{Y^n}(\mathcal{I}_{g_n}) \leq \delta\right\}. \quad (28)$$

The set for $\delta = 0$ is called *secure output set* and denoted $\mathcal{T}_0$.

### D. $(\epsilon, \delta)$-Secure Capacity

We are now prepared to derive the first fundamental result - the $(\epsilon, \delta)$-Secure Capacity. This capacity will make use of the following definition,

$$J(R|\mathbf{X}) := \limsup_{n \to \infty} \Pr\left\{\frac{1}{n} i(X^n; Z^n) \leq R\right\}$$
$$= \limsup_{n \to \infty} \Pr\left\{\frac{1}{n} \log \frac{Q^n(Z^n|X^n)}{p_{Z^n}(Z^n)} \leq R\right\}.$$

The proof is the general $\epsilon$-capacity proof given by Han[6], [7], with the restriction to the secure input set.

*Theorem 2.1 ($(\epsilon, \delta)$-Secure Capacity):* The $(\epsilon, \delta)$-secure capacity $C(\epsilon, \delta|\mathbf{W}, \mathbf{g}, \mathbf{A})$ of a stego-channel $(\mathbf{W}, \mathbf{g}, \mathbf{A})$ is given by,

$$C(\epsilon, \delta|\mathbf{W}, \mathbf{g}, \mathbf{A}) = \sup_{\mathbf{X} \in \mathcal{S}_\delta} \sup\{R : J(R|\mathbf{X}) \leq \epsilon\}, \quad (29)$$

for any $0 \leq \epsilon < 1$ and $0 \leq \delta < 1$.

*Proof:* This proof is based on [6], [7]. Let $C = \sup_{\mathbf{X} \in \mathcal{S}_\delta} \sup\{R : J(R|\mathbf{X}) \leq \epsilon\}$, and $Q^n = A^n \circ W^n$.

**Achievability**: Choose any $\epsilon \geq 0$ and $\delta > 0$.

Let $R = C - 3\gamma$, for any $\gamma > 0$. By the definition of $C$ we have that there exists an $\mathbf{X} \in \mathcal{S}_\delta$ such that,

$$\sup\{R : J(R|\mathbf{X}) \leq \epsilon\} \geq C - \gamma = R + 2\gamma. \quad (30)$$

Similarly we may find an $R' > R + \gamma$ such that $J(R'|\mathbf{X}) \leq \epsilon$. As $J(R|\mathbf{X})$ is monotonically increasing,

$$J(R + \gamma|\mathbf{X}) \leq \epsilon. \quad (31)$$

Next by letting $M_n = e^{nR}$ we have that,

$$\liminf_{n \to \infty} \frac{1}{n} \log M_n \geq R.$$

Using Feinstein's Lemma[11] we have that there exists an $(n, M_n, \epsilon_n)$-code with,

$$\epsilon_n \leq \Pr\left\{\frac{1}{n} \log \frac{Q^n(Z^n|X^n)}{p_{Z^n}(Z^n)} \leq \frac{1}{n} \log M_n + \gamma\right\} + e^{-n\gamma}. \quad (32)$$

As $\frac{1}{n} \log M_n = R$ for all $n$ we have,

$$\epsilon_n \leq \Pr\left\{\frac{1}{n} \log \frac{Q^n(Z^n|X^n)}{p_{Z^n}(Z^n)} \leq R + \gamma\right\} + e^{-n\gamma}. \quad (33)$$

Taking the $\limsup$ of each side we have,

$$\limsup_{n \to \infty} \epsilon_n \leq J(R + \gamma|\mathbf{X}), \quad (34)$$

with $J(R + \gamma|\mathbf{X}) \leq \epsilon$ shows that $\limsup_{n \to \infty} \epsilon_n \leq \epsilon$.
Finally since $\mathbf{X} \in \mathcal{S}_\delta$ we have that,

$$\limsup_{n \to \infty} p_{Z^n}(\mathcal{I}_{g_n}) \leq \delta. \quad (35)$$

**Converse**: Let $R > C$, and choose $\gamma > 0$ such that $R - 2\gamma > C$. Assume that $R$ is $(\epsilon, \delta)$-achievable, so there exists an $(n, M_n, \epsilon_n, \delta_n)$-code such that,

$$\liminf_{n \to \infty} \frac{1}{n} \log M_n \geq R, \quad (36)$$

$$\limsup_{n \to \infty} \epsilon_n \leq \epsilon, \quad (37)$$

and

$$\limsup_{n \to \infty} \delta_n \leq \delta. \quad (38)$$

Let $\mathbf{X} = \{X^n\}_{n=1}^{\infty}$ where each $X^n$ is a uniform distribution over codewords $\mathcal{C}_n$, and let $\mathbf{Z}$ be the corresponding channel output. Since $R - 2\gamma > C \geq \sup\{R : J(R|\mathbf{X}) \leq \epsilon\}$,

$$J(R - 2\gamma|\mathbf{X}) > \epsilon. \quad (39)$$

The Feinstein Dual [6], [7] states that for a uniformly distributed input $X^n$ over a $(n, M_n, \epsilon_n)$-code and output $Z^n$ corresponding to channel $\mathbf{Q}$, the following holds for all $n$,

$$\epsilon_n \geq \Pr\left\{\frac{1}{n} \log \frac{Q^n(Z^n|X^n)}{p_{Z^n}(Z^n)} \leq \frac{1}{n} \log M_n - \gamma\right\} - e^{-n\gamma} \quad (40)$$

Using the property of $\liminf$ we have that for all $n > n_0$ that,

$$\frac{1}{n} \log M_n \geq R - \gamma. \quad (41)$$

For $n > n_0$ we have,

$$\epsilon_n \geq \Pr\left\{\frac{1}{n} \log \frac{Q^n(Z^n|X^n)}{p_{Z^n}(Z^n)} \leq R - 2\gamma\right\} - e^{-n\gamma}. \quad (42)$$

Taking the $\limsup$ of both sides, and considering (39), we see that,

$$\limsup_{n \to \infty} \epsilon_n > \epsilon. \quad (43)$$

∎

A fundamental assumption in the above proof is that the encoder has a knowledge of the detection function. From a steganalysis perspective this allows one to determine the "worst-case scenario" for the amount of information that may be sent through a channel.

### E. Secure Capacity

The next result deals with a special case of $(\epsilon, \delta)$-secure capacity, namely the one where $\epsilon = \delta = 0$. The secure capacity is the maximum amount of information that may be sent over a channel with arbitrarily small probabilities of error and detection.

The four potential formulations for our model are shown in Figure 3. The capacity of the stego-channel $(\mathbf{W}, \mathbf{g}, \mathbf{A})$ is shown in Theorem 2.2 to follow and specialized to the other cases in Theorems 2.3, 2.4 and 2.5.

The results of these capacities are summarized in Table II.

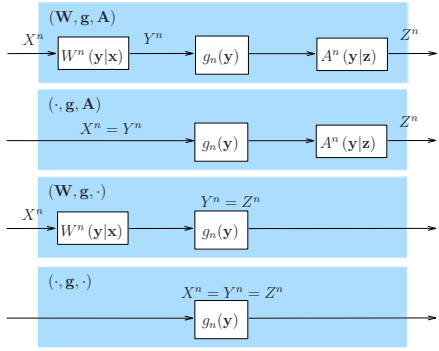

Fig. 3. Stegochannels

TABLE II
SECURE CAPACITY FORMULAS

| Secure Capacity | Noise | Attack | Thm. |
|---|---|---|---|
| $C(\mathbf{W}, \mathbf{g}, \mathbf{A}) = \sup_{\mathbf{X} \in \mathcal{S}_0} \underline{I}(\mathbf{X}; \mathbf{Z})$ | W | A | 2.2 |
| $C(\cdot, \mathbf{g}, \mathbf{A}) = \sup_{\mathbf{Y} \in \mathcal{T}_0} \underline{I}(\mathbf{Y}; \mathbf{Z})$ | Noiseless | A | 2.3 |
| $C(\mathbf{W}, \mathbf{g}) = \sup_{\mathbf{X} \in \mathcal{S}_0} \underline{I}(\mathbf{X}; \mathbf{Y})$ | W | Passive | 2.4 |
| $C(\cdot, \mathbf{g}) = \sup_{\mathbf{Y} \in \mathcal{T}_0} \underline{H}(\mathbf{Y})$ | Noiseless | Passive | 2.5 |

*Theorem 2.2 (Secure Capacity):* The secure capacity $C(\mathbf{W}, \mathbf{g}, \mathbf{A})$ of a stego-channel $(\mathbf{W}, \mathbf{g}, \mathbf{A})$ is given by,

$$C(\mathbf{W}, \mathbf{g}, \mathbf{A}) = \sup_{\mathbf{X} \in \mathcal{S}_0} \underline{I}(\mathbf{X}; \mathbf{Z}). \quad (44)$$

*Proof:* We apply Theorem 2.1 with $\epsilon = 0$ and $\delta = 0$. This gives,

$$C(\mathbf{W}, \mathbf{g}, \mathbf{A})$$
$$= C(0, 0 | \mathbf{W}, \mathbf{g}, \mathbf{A}) \quad (45a)$$
$$= \sup_{\mathbf{X} \in \mathcal{S}_0} \sup \{R : J(R|\mathbf{X}) \leq 0\} \quad (45b)$$
$$= \sup_{\mathbf{X} \in \mathcal{S}_0} \sup \left[ R : \limsup_{n \to \infty} \Pr \left\{ \frac{1}{n} i(X^n; Z^n) \leq R \right\} \leq 0 \right] \quad (45c)$$
$$= \sup_{\mathbf{X} \in \mathcal{S}_0} \underline{I}(\mathbf{X}; \mathbf{Z}) \quad (45d)$$

Here the last line is due to the definition of p-$\liminf$. ■

*Theorem 2.3 (Noiseless Encoder, Active Adversary):* The secure capacity of a stego-channel, $(\cdot, \mathbf{g}, \mathbf{A})$, with a noiseless-encoder and active adversary, denoted $C(\cdot, \mathbf{g}, \mathbf{A})$, is given by,

$$C(\cdot, \mathbf{g}, \mathbf{A}) = \sup_{\mathbf{Y} \in \mathcal{T}_0} \underline{I}(\mathbf{Y}; \mathbf{Z}). \quad (46)$$

*Proof:* Apply Theorem 2.2 with $\mathbf{X} = \mathbf{Y}$ and $\mathcal{S}_0 = \mathcal{T}_0$. ■

*Theorem 2.4 (Passive Adversary):* The secure channel capacity with a passive adversary, denoted $C(\mathbf{W}, \mathbf{g})$ of a stego-channel $(\mathbf{W}, \mathbf{g}, \cdot)$ is given by,

$$C(\mathbf{W}, \mathbf{g}) = \sup_{\mathbf{X} \in \mathcal{S}_0} \underline{I}(\mathbf{X}; \mathbf{Y}). \quad (47)$$

*Proof:* Since the adversary is passive, we have that $\mathbf{Z} = \mathbf{Y}$. ■

*Theorem 2.5 (Noiseless Encoder, Passive Adversary):* The secure capacity of a stego-channel $(\cdot, \mathbf{g}, \cdot)$, with a noiseless-encoder and passive adversary, denoted $C(\cdot, \mathbf{g})$, is given by,

$$C(\cdot, \mathbf{g}) = \sup_{\mathbf{X} \in \mathcal{S}_0} \underline{I}(\mathbf{X}; \mathbf{Y}). \quad (48)$$

*Proof:* Since the adversary is passive, we have that $\mathbf{Z} = \mathbf{Y}$, and since there is no encoder noise we have that $\mathbf{X} = \mathbf{Y}$ and $\mathcal{S}_0 = \mathcal{T}_0$. ■

### F. Strong Converse

A stego-channel $(\mathbf{W}, \mathbf{g}, \mathbf{A})$ is said to satisfy the $\epsilon$-strong converse property if for any $R > C(0, \delta | \mathbf{W}, \mathbf{g}, \mathbf{A})$, every $(n, M_n, \epsilon_n, \delta_n)$-code with,

$$\liminf_{n \to \infty} \frac{1}{n} \log M_n \geq R,$$

and

$$\limsup_{n \to \infty} \delta_n \leq \delta,$$

we have,

$$\lim_{n \to \infty} \epsilon_n = 1.$$

If a channel satisfies the $\epsilon$-strong converse,

$$C(\epsilon, \delta | \mathbf{W}, \mathbf{g}, \mathbf{A}) = C(0, \delta | \mathbf{W}, \mathbf{g}, \mathbf{A}), \quad (49)$$

for any $\epsilon \in [0, 1)$.

*Theorem 2.6 ($\epsilon$-Strong Converse):* A stego-channel $(\mathbf{W}, \mathbf{g}, \mathbf{A})$ satisfies the $\epsilon$-strong converse property (for a fixed $\delta$) if and only if,

$$\sup_{\mathbf{X} \in \mathcal{S}_\delta} \underline{I}(\mathbf{X}; \mathbf{Z}) = \sup_{\mathbf{X} \in \mathcal{S}_\delta} \overline{I}(\mathbf{X}; \mathbf{Z}). \quad (50)$$

This proof is essentially the $\epsilon$-strong converse[6], [7] with a restriction to the secure input set. See details in Appendix A

### G. Bounds

We now derive a number of useful bounds on the spectral-entropy of an output sequence in relation to the permissible set. These bounds will then be used to prove general bounds for steganographic systems and see further application in Chapter III.

*Theorem 2.7 (Spectral inf-entropy bound):* For a discrete $\mathbf{g} = \{\mathcal{P}_n\}_{n=1}^{\infty}$ with corresponding secure output set $\mathcal{T}_0$,

$$\sup_{\mathbf{Y} \in \mathcal{T}_0} \underline{H}(\mathbf{Y}) = \liminf_{n \to \infty} \frac{1}{n} \log |\mathcal{P}_n| \quad (51)$$

See Appendix B for proof.

*Theorem 2.8 (Spectral sup-entropy bound):* For discrete $\mathbf{g} = \{\mathcal{P}_n\}_{n=1}^{\infty}$ with corresponding secure output set $\mathcal{T}_0$,

$$\sup_{\mathbf{Y} \in \mathcal{T}_0} \overline{H}(\mathbf{Y}) = \limsup_{n \to \infty} \frac{1}{n} \log |\mathcal{P}_n| \quad (52)$$

See Appendix C for proof.

### H. Capacity Bounds

This section present a number of fundamental bounds on the secure capacity of a stego-channel based on the properties of that channel.

We make use of the following lemma,

*Lemma 2.1:* For a stego-channel $(\mathbf{W}, \mathbf{g}, \mathbf{A})$ the following hold,

$$\underline{I}(\mathbf{X}; \mathbf{Z}) \leq \underline{I}(\mathbf{X}; \mathbf{Y}), \qquad (53)$$
$$\underline{I}(\mathbf{X}; \mathbf{Z}) \leq \underline{I}(\mathbf{Y}; \mathbf{Z}). \qquad (54)$$

*Proof:* We note that the general distributions form a Markov chain, $\mathbf{X} \to \mathbf{Y} \to \mathbf{Z}^1$. A property of the inf-information rate[7] is,

$$\underline{I}(\mathbf{X}; \mathbf{Z}) \leq \underline{I}(\mathbf{X}; \mathbf{Y}), \qquad (55)$$

when $\mathbf{X} \to \mathbf{Y} \to \mathbf{Z}$.

Since $\mathbf{X} \to \mathbf{Y} \to \mathbf{Z}$ implies $\mathbf{Z} \to \mathbf{Y} \to \mathbf{X}$ we also have,

$$\underline{I}(\mathbf{X}; \mathbf{Z}) \leq \underline{I}(\mathbf{Y}; \mathbf{Z}). \qquad (56)$$

∎

The first capacity bound gives an upperbound based on the sup-entropy of the secure input set.

*Theorem 2.9 (Input Sup-Entropy Bound):* For a stego-channel $(\mathbf{W}, \mathbf{g}, \mathbf{A})$ the secure capacity is bounded as,

$$C(\mathbf{W}, \mathbf{g}, \mathbf{A}) \leq \sup_{\mathbf{X} \in \mathcal{S}_0} \overline{H}(\mathbf{X}) \qquad (57)$$

*Proof:* Using (21) and the property that $\overline{H}(\mathbf{X}|\mathbf{Z}) \geq 0$ we have,

$$C(\mathbf{W}, \mathbf{g}, \mathbf{A}) \stackrel{(T2.2)}{=} \sup_{\mathbf{X} \in \mathcal{S}_0} \underline{I}(\mathbf{X}; \mathbf{Z})$$
$$\stackrel{(21)}{\leq} \sup_{\mathbf{X} \in \mathcal{S}_0} \{\overline{H}(\mathbf{X}) - \overline{H}(\mathbf{X}|\mathbf{Z})\}$$
$$\leq \sup_{\mathbf{X} \in \mathcal{S}_0} \overline{H}(\mathbf{X})$$

∎

The next theorem gives two upper bounds on the capacity based on the sup-entropy of the secure input and output sets.

*Theorem 2.10 (Output Sup-Entropy Bounds):* For a stego-channel $(\mathbf{W}, \mathbf{g}, \mathbf{A})$ the secure capacity is bounded as,

$$C(\mathbf{W}, \mathbf{g}, \mathbf{A}) \leq \sup_{\mathbf{X} \in \mathcal{S}_0} \overline{H}(\mathbf{Y}) \qquad (59a)$$
$$\leq \sup_{\mathbf{Y} \in \mathcal{T}_0} \overline{H}(\mathbf{Y}) \qquad (59b)$$

*Proof:* Using (21) and the property that $\overline{H}(\mathbf{Z}|\mathbf{X}) \geq 0$ we have,

$$C(\mathbf{W}, \mathbf{g}, \mathbf{A}) = \sup_{\mathbf{X} \in \mathcal{S}_0} \underline{I}(\mathbf{X}; \mathbf{Z})$$
$$\stackrel{(L2.1)}{\leq} \sup_{\mathbf{X} \in \mathcal{S}_0} \underline{I}(\mathbf{X}; \mathbf{Y})$$
$$\stackrel{(21)}{\leq} \sup_{\mathbf{X} \in \mathcal{S}_0} \{\overline{H}(\mathbf{Y}) - \overline{H}(\mathbf{Y}|\mathbf{X})\}$$
$$\leq \sup_{\mathbf{X} \in \mathcal{S}_0} \overline{H}(\mathbf{Y})$$
$$\leq \sup_{\mathbf{Y} \in \mathcal{T}_0} \overline{H}(\mathbf{Y})$$

---

[1] $\mathbf{X} \to \mathbf{Y} \to \mathbf{Z}$ is said to hold when for all $n$, $X^n$ and $Z^n$ are conditionally independent given $Y^n$.

Here the final line follows since if $\mathbf{X} \in \mathcal{S}_0$ and $\mathbf{X} \stackrel{\mathbf{W}}{\to} \mathbf{Y}$ then $\mathbf{Y} \in \mathcal{T}_0$. ∎

The next corollary specializes the above theorem when the permissible set is finite.

*Corallary 2.1 (Discrete Permissible Set Bound):* For a given discrete stego-channel $(\mathbf{W}, \mathbf{g}, \mathbf{A}) = \{(W^n, \mathcal{P}_{g_n}, A^n)\}_{n=1}^{\infty}$ the secure capacity is bounded from above as,

$$C(\mathbf{W}, \mathbf{g}, \mathbf{A}) \leq \limsup_{n \to \infty} \frac{1}{n} \log |\mathcal{P}_{g_n}| \qquad (61)$$

*Proof:* Combining Theorem 2.8 and line (59b) of Theorem 2.10 gives the desired result. ∎

The next theorem provides an intuitive result dealing with the capacity of two stego-channels having related steganalyzers.

*Theorem 2.11 (Permissible Set Relation):* For two stego-channels, $(\mathbf{W}, \mathbf{g}, \mathbf{A})$ and $(\mathbf{W}, \mathbf{v}, \mathbf{A})$ if $\mathcal{P}_{g_n} \subseteq \mathcal{P}_{v_n}$ for all but finitely many $n$, then,

$$C(\mathbf{W}, \mathbf{g}, \mathbf{A}) \leq C(\mathbf{W}, \mathbf{v}, \mathbf{A}). \qquad (62)$$

*Proof:* Let $\{f_n\}_{n=1}^{\infty}$ and $\{\phi_n\}_{n=1}^{\infty}$ be a sequence of encoding and decoding functions that achieves $C(\mathbf{W}, \mathbf{g}, \mathbf{A})$. Such a sequence exists by the definition of secure capacity. The following definitions will be used for $i = 1, \ldots, M_n$,

$$\mathbf{u}_i = f_n(i),$$
$$\mathcal{D}_i = \phi_n^{-1}(\{i\}).$$

The probability of error for this sequence is given by (12),

$$\epsilon_n = \frac{1}{M_n} \sum_{i=1}^{M_n} Q^n (\mathcal{D}_i^c | \mathbf{u}_i),$$

where $Q^n = A^n \circ W^n$.

This value is independent of the permissible sets and if $\epsilon_n \to 0$ for the stego-channel $(\mathbf{W}, \mathbf{g}, \mathbf{A})$ then it also goes to zero for $(\mathbf{W}, \mathbf{v}, \mathbf{A})$.

Next we know that the probability of detection for $(\mathbf{W}, \mathbf{g}, \mathbf{A})$ is given by (13),

$$\delta_n^{\mathbf{g}} = \frac{1}{M_n} \sum_{i=1}^{M_n} W^n (\mathcal{I}_{g_n} | \mathbf{u}_i),$$

and that $\delta_n^{\mathbf{g}} \to 0$.

Since $\mathcal{P}_{g_n} \subseteq \mathcal{P}_{v_n}$ for all $n > N$, we have that, $\mathcal{I}_{g_n} \supseteq \mathcal{I}_{v_n}$ if $n > N$ and,

$$W^n (\mathcal{I}_{g_n} | \mathbf{x}) \geq W^n (\mathcal{I}_{v_n} | \mathbf{x}), \qquad \forall n > N, \mathbf{x} \in \mathcal{X}^n. \quad (63)$$

Using this, we may bound the probability of detection for $(\mathbf{W}, \mathbf{v}, \mathbf{A})$ and $n > N$ as,

$$\delta_n^{\mathbf{v}} = \frac{1}{M_n} \sum_{i=1}^{M_n} W^n (\mathcal{I}_{v_n} | \mathbf{u}_i)$$
$$\stackrel{(63)}{\leq} \frac{1}{M_n} \sum_{i=1}^{M_n} W^n (\mathcal{I}_{g_n} | \mathbf{u}_i)$$
$$= \delta_n^{\mathbf{g}}$$

Since $\delta_n^{\mathbf{g}} \to 0$ we see that $\delta_n^{\mathbf{v}} \to 0$ as well. ∎

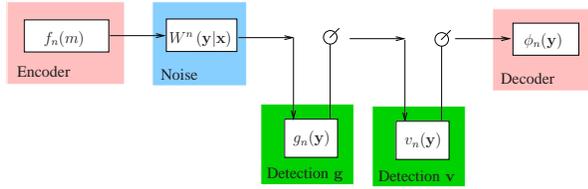

Fig. 4. Composite steganalyzer

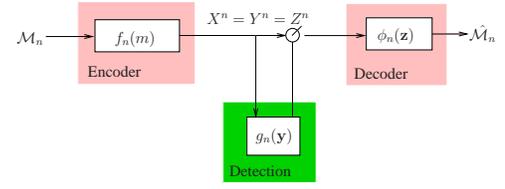

Fig. 6. Noiseless Stego-Channel

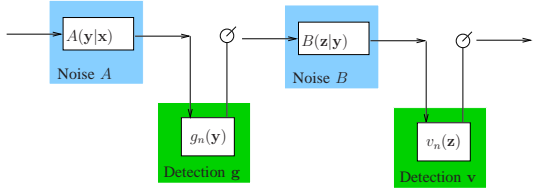

Fig. 5. Two Noise Channel

## I. Applications

*1) Composite steganalyzers:* The final theorem of the previous section is intuitively pleasing and leads to some immediate results. An example of this is the composite steganalyzer pictured in Figure 4.

In this system, two steganalyzers, $\mathbf{g}$ and $\mathbf{v}$ are used sequentially on the corrupted stego-signal. If either of these steganalyzers are triggered, the message is considered steganographic. We will denote the composite stego-channel of this system as $(\mathbf{W}, \mathbf{h}, \mathbf{A})$.

As one would expect, the capacity of the composite channel, $C(\mathbf{W}, \mathbf{h}, \mathbf{A})$, is smaller than either $C(\mathbf{W}, \mathbf{g}, \mathbf{A})$ or $C(\mathbf{W}, \mathbf{v}, \mathbf{A})$. This is shown in the next theorem.

*Theorem 2.12 (Composite Stego-Channel):* For a composite stego-channel $(\mathbf{W}, \mathbf{h}, \mathbf{A})$ defined by $\mathbf{g}$ and $\mathbf{v}$, the following inequality holds,

$$C(\mathbf{W}, \mathbf{h}, \mathbf{A}) \leq \min\{C(\mathbf{W}, \mathbf{g}, \mathbf{A}), C(\mathbf{W}, \mathbf{v}, \mathbf{A})\}. \quad (65)$$

*Proof:* We first show that $C(\mathbf{W}, \mathbf{h}, \mathbf{A}) \leq C(\mathbf{W}, \mathbf{g}, \mathbf{A})$.

The permissible set of the composite is equal to the intersection of the base detection functions,

$$\mathcal{P}_{h_n} = \mathcal{P}_{g_n} \cap \mathcal{P}_{v_n}, \quad \forall n, \quad (66)$$

thus we have that $\mathcal{P}_{h_n} \subseteq \mathcal{P}_{g_n}$ and we may apply Theorem 2.11 to state,

$$C(\mathbf{W}, \mathbf{h}, \mathbf{A}) \leq C(\mathbf{W}, \mathbf{g}, \mathbf{A}).$$

The above argument may be applied using $\mathcal{P}_{h_n} \subseteq \mathcal{P}_{v_n}$ to show $C(\mathbf{W}, \mathbf{h}, \mathbf{A}) \leq C(\mathbf{W}, \mathbf{v}, \mathbf{A})$. ∎

*2) Two Noise Systems:* We briefly present and discuss an interesting case that is somewhat counter-intuitive. Consider the channel shown in Figure 5. In this case there is distortion $A$ after the encoder and a second distortion $B$ before the second steganalyzer. In the previous section it was shown that in the composite steganalyzer the addition of a second steganalyzer (Figure 5) lowers the capacity of the stego-channel. A surprising result for the two noise system is that this may not be the case. In fact, the addition of a second distortion may increase the capacity of a stego-channel!

To see this, consider the two steganalyzers $\mathbf{g}$ and $\mathbf{v}$. Assume that $\mathbf{g}$ classifies signals with positive means as steganographic, while $\mathbf{v}$ classifies signals with negative means as steganographic. If these detection functions were in series, the permissible set (of the composite detection function) is empty. This is because a signal cannot have a positive and negative mean. Now consider a specific, deterministic distortion $B^n(-\mathbf{y}|\mathbf{y}) = 1$. Now we may send any signal we wish, as long as its mean is positive. So in some instances, it is possible for the addition of a distortion to actually increase the capacity.

## III. NOISELESS CHANNELS

This section investigates the capacity of the noiseless stego-channel shown in Figure 6. In this system there is no encoder-noise and the adversary is passive. This means that not only does the decoder receive exactly what the encoder sends, but the steganalyzer does as well.

This section finds the secure capacity of this system, and then derives a number of intuitive bounds relating to this capacity.

### A. Secure Noiseless Capacity

*Theorem 3.1 (Secure Noiseless Capacity):* For a discrete noiseless channel $(\cdot, \mathbf{g}, \cdot)$ the secure capacity is given by,

$$C(\cdot, \mathbf{g}) = \liminf_{n\to\infty} \frac{1}{n} \log |\mathcal{P}_{g_n}| \quad (67)$$

*Proof:* The proof follows directly from Theorem 2.5 and Theorem 2.7. ∎

*Example 2 (Capacity of the Sum Steganalyzer):* We now use this result to find the secure noiseless capacity of the sum steganalyzer of Example 1. The size of the permissible set for $n$ is equal to the number of different ways we may arrange up to $\lfloor n/2 \rfloor$ 1s into $n$ positions.

$$|\mathcal{P}_{g_n}| = \sum_{i:0\leq i \leq \lfloor \frac{n}{2} \rfloor} \binom{n}{i}. \quad (68)$$

For $n$ even $|\mathcal{P}_{g_n}| = 2^{n-1} + \frac{1}{2}\binom{n}{n/2}$ and for $n$ odd, $|\mathcal{P}_{g_n}| = 2^{n-1}$. Applying the noiseless Theorem,

$$C(\cdot, \mathbf{g}) = \liminf_{n\to\infty} \frac{1}{n} \log |\mathcal{P}_{g_n}| = \lim_{n\to\infty} \frac{1}{n} \log 2^{n-1}$$
$$= 1 \text{bit/use}. \quad (69a)$$

## B. $\epsilon$-Strong Converse for Noiseless Channels

We now present a fundamental result for discrete noiseless channels regarding the $\epsilon$-strong converse property. It gives the necessary and sufficient conditions for a noiseless stego-channel to satisfy the $\epsilon$-strong converse property.

*Theorem 3.2 (Noiseless $\epsilon$-Strong Converse):* A discrete noiseless stego-channel $(\cdot, \mathbf{g}, \cdot)$ satisfies the $\epsilon$-strong converse property if and only if,

$$C(\cdot, \mathbf{g}) = \lim_{n \to \infty} \frac{1}{n} \log |\mathcal{P}_{g_n}|. \tag{70}$$

*Proof:* Since the channel is noiseless, $\mathbf{X} = \mathbf{Y} = \mathbf{Z}$ we have,

$$\sup_{\mathbf{X} \in \mathcal{S}_0} \underline{I}(\mathbf{X}; \mathbf{Z}) = \sup_{\mathbf{Y} \in \mathcal{T}_0} \underline{H}(\mathbf{Y}), \tag{71}$$

$$\sup_{\mathbf{X} \in \mathcal{S}_0} \overline{I}(\mathbf{X}; \mathbf{Z}) = \sup_{\mathbf{Y} \in \mathcal{T}_0} \overline{H}(\mathbf{Y}). \tag{72}$$

First assume that the stego-channel satisfies the $\epsilon$-strong converse property. This gives,

$$\sup_{\mathbf{Y} \in \mathcal{T}_0} \underline{H}(\mathbf{Y}) \stackrel{(71)}{=} \sup_{\mathbf{X} \in \mathcal{S}_0} \underline{I}(\mathbf{X}; \mathbf{Z}) \tag{73a}$$

$$\stackrel{(T2.6)}{=} \sup_{\mathbf{X} \in \mathcal{S}_0} \overline{I}(\mathbf{X}; \mathbf{Z}) \tag{73b}$$

$$\stackrel{(72)}{=} \sup_{\mathbf{Y} \in \mathcal{T}_0} \overline{H}(\mathbf{Y}) \tag{73c}$$

The capacity is then,

$$C(\cdot, \mathbf{g}) \stackrel{(T2.5)}{=} \sup_{\mathbf{Y} \in \mathcal{T}_0} \underline{H}(\mathbf{Y})$$

$$\stackrel{(T2.7)}{=} \liminf_{n \to \infty} \frac{1}{n} \log |\mathcal{P}_{g_n}|$$

$$\stackrel{(73c)}{=} \sup_{\mathbf{Y} \in \mathcal{T}_0} \overline{H}(\mathbf{Y})$$

$$\stackrel{(T2.8)}{=} \limsup_{n \to \infty} \frac{1}{n} \log |\mathcal{P}_{g_n}|$$

$$= \lim_{n \to \infty} \frac{1}{n} \log |\mathcal{P}_{g_n}|$$

Here the final line results as the $\liminf$ and $\limsup$ coincide.

For the other direction assume that $C(\cdot, \mathbf{g}) = \lim_{n \to \infty} \frac{1}{n} \log |\mathcal{P}_{g_n}|$ which gives,

$$C(\cdot, \mathbf{g}) = \sup_{\mathbf{X} \in \mathcal{S}_0} \underline{I}(\mathbf{X}; \mathbf{Z})$$

$$\stackrel{(T2.5)}{=} \sup_{\mathbf{Y} \in \mathcal{T}_0} \underline{H}(\mathbf{Y})$$

$$= \lim_{n \to \infty} \frac{1}{n} \log |\mathcal{P}_{g_n}|$$

$$= \limsup_{n \to \infty} \frac{1}{n} \log |\mathcal{P}_{g_n}|$$

$$\stackrel{(T2.8)}{=} \sup_{\mathbf{Y} \in \mathcal{T}_0} \overline{H}(\mathbf{Y})$$

$$\stackrel{(72)}{=} \sup_{\mathbf{X} \in \mathcal{S}_0} \overline{I}(\mathbf{X}; \mathbf{Z})$$

Thus, $\sup_{\mathbf{X} \in \mathcal{S}_0} \underline{I}(\mathbf{X}; \mathbf{Z}) = \sup_{\mathbf{X} \in \mathcal{S}_0} \overline{I}(\mathbf{X}; \mathbf{Z})$ and by Theorem 2.6 the stego-channel satisfies the $\epsilon$-strong-converse property. ∎

*Example 3 (Sum Steganalyzer):* We now determine if the sum steganalyzer satisfies the $\epsilon$-strong converse.

From Example 2 the size of the permissible set is,

$$|\mathcal{P}_{g_n}| = \begin{cases} 2^{n-1} + \frac{1}{2}\begin{pmatrix} n \\ n/2 \end{pmatrix}, & \text{for even } n \\ 2^{n-1}, & \text{for odd } n \end{cases} \tag{75}$$

We will make use of Stirling's approximation,

$$n! = \sqrt{2\pi} n^{n+\frac{1}{2}} e^{-n+\lambda_n}, \tag{76}$$

where $1/(12n+1) < \lambda_n < 1/(12n)$.

For $n$ even,

$$|\mathcal{P}_{g_n}| = 2^{n-1} + \frac{1}{2} \frac{n!}{(n - \frac{1}{2}n)!(\frac{1}{2}n)!} \tag{77}$$

$$= 2^{n-1} + \frac{1}{2} \frac{\sqrt{2\pi} n^{n+\frac{1}{2}} e^{-n+\lambda_n}}{\left( \sqrt{2\pi}(n/2)^{\frac{n}{2}+\frac{1}{2}} e^{-\frac{n}{2}+\lambda_{n/2}} \right)^2} \tag{78}$$

$$\leq 2^{n-1} \left( 1 + \frac{2e}{\sqrt{2\pi n}} \right) \tag{79}$$

This gives,

$$\limsup_{n \to \infty} \frac{1}{n} \log |\mathcal{P}_{g_n}|$$

$$\leq \limsup_{n \to \infty} \frac{1}{n} \log \left( 2^{n-1} \left( 1 + \frac{2e}{\sqrt{2\pi n}} \right) \right) \tag{80}$$

$$= 1 \tag{81}$$

This shows,

$$\liminf_{n \to \infty} \frac{1}{n} \log |\mathcal{P}_{g_n}| = 1 \geq \limsup_{n \to \infty} \frac{1}{n} \log |\mathcal{P}_{g_n}|. \tag{82}$$

Since the liminf and limsup coincide, the limit is indeed a true one and this stego-channel satisfies the $\epsilon$-strong converse.

## C. Properties of the Noiseless DMSC

In this section we briefly investigate the secure capacity of the discrete memoryless stego-channel (cf. I-F2).

*Theorem 3.3 (Noiseless DMSC Secure Capacity):* For the stego-channel $(\cdot, \mathbf{g}, \cdot)$ with $\mathbf{g} = \{g\}$, the secure capacity is given by,

$$C(\cdot, \mathbf{g}) = \log |\mathcal{P}_g|, \tag{83}$$

and furthermore this stego-channel satisfies the strong converse.

*Proof:* As the channel is noiseless and the input alphabet is finite we may use Theorem 3.1,

$$C(\cdot, \mathbf{g}) = \liminf_{n \to \infty} \frac{1}{n} \log |\mathcal{P}_{g_n}|. \tag{84}$$

Note that by (7) we have for all $n$,

$$\frac{1}{n} \log |\mathcal{P}_{g_n}| = \frac{1}{n} \log \left| \underbrace{\mathcal{P}_g \times \mathcal{P}_g \times \cdots \times \mathcal{P}_g}_{n} \right|$$

$$= \frac{1}{n} \log |\mathcal{P}_g|^n$$

$$= \log |\mathcal{P}_g|.$$

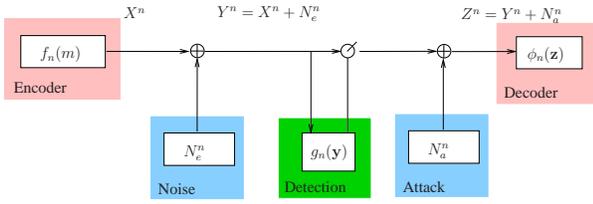

Fig. 7. Additive Noise Channel Active Adversary

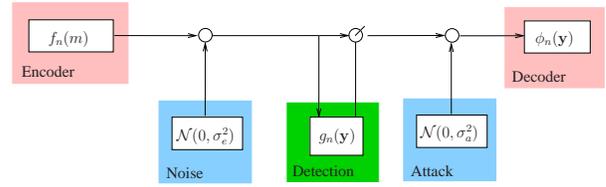

Fig. 8. AWGN Channel Active Adversary

Thus,
$$C(\cdot, \mathbf{g}) = \log |\mathcal{P}_g|. \quad (85)$$

We also have that
$$C(\cdot, \mathbf{g}) = \liminf_{n \to \infty} \frac{1}{n} \log |\mathcal{P}_{g_n}| = \log |\mathcal{P}_g| = \lim_{n \to \infty} \frac{1}{n} \log |\mathcal{P}_{g_n}|, \quad (86)$$
thus by Theorem 3.2 the stego-channel satisfies the strong converse. ∎

## IV. ADDITIVE NOISE STEGO-CHANNELS

In this section we evaluate the capacity of particular stego-channel, shown in Figure 7. In this channel, both the encoder-noise and attack-noise are additive and independent from the channel input.

### A. Additive Noise

Denote the sum of two general sequences $\mathbf{X} = \{X^n = (X_1^{(n)}, \ldots, X_n^{(n)})\}_{n=1}^{\infty}$, and $\mathbf{Y} = \{Y^n = (Y_1^{(n)}, \ldots, Y_n^{(n)})\}_{n=1}^{\infty}$ as,

$$\mathbf{X} + \mathbf{Y} := \{X^n + Y^n = (X_1^{(n)} + Y_1^{(n)}, \ldots, X_n^{(n)} + Y_n^{(n)})\}_{n=1}^{\infty}. \quad (87)$$

Letting the encoder-noise be denoted as $\mathbf{N}_e = \{N_e^n\}_{n=1}^{\infty}$ and the attack-noise denoted as $\mathbf{N}_a = \{N_a^n\}_{n=1}^{\infty}$ we have the following relations,
$$\mathbf{Y} = \mathbf{X} + \mathbf{N}_e$$
$$\mathbf{Z} = \mathbf{Y} + \mathbf{N}_a = \mathbf{X} + \mathbf{N}_e + \mathbf{N}_a = \mathbf{X} + \mathbf{N}$$
where $\mathbf{N} = \{N^n\}_{n=1}^{\infty} = \mathbf{N}_e + \mathbf{N}_a$.

As noises are independent from the stego-signal, we may use the following simplifications,
$$p_{Z^n|X^n}(X^n + N^n | X^n) = p_{N^n}(N^n),$$
leading to the following simplifications in spectral-entropies,
$$\underline{H}(\mathbf{Z}|\mathbf{X}) = \underline{H}(\mathbf{N}), \quad (88)$$
$$\overline{H}(\mathbf{Z}|\mathbf{X}) = \overline{H}(\mathbf{N}). \quad (89)$$

We now use these simplifications to present a useful capacity result for additive noise channels.

*Theorem 4.1:* For additive noise stego-channel defined with $\mathbf{N}_e + \mathbf{N}_a = \mathbf{N}$, if $\mathbf{N}$ satisfies the strong converse (i.e. $\underline{H}(\mathbf{N}) = \overline{H}(\mathbf{N})$) then the capacity is,
$$C(\mathbf{W}, \mathbf{g}, \mathbf{A}) = \sup_{\mathbf{X} \in \mathcal{S}_0} \{\underline{H}(\mathbf{Z})\} - \underline{H}(\mathbf{N}) \quad (90)$$

*Proof:* First we find a lower bound as,
$$C(\mathbf{W}, \mathbf{g}, \mathbf{A}) \stackrel{(22)}{\geq} \sup_{\mathbf{X} \in \mathcal{S}_0} \{\underline{H}(\mathbf{Z}) - \overline{H}(\mathbf{Z}|\mathbf{X})\} \quad (91)$$
$$\stackrel{(88)}{=} \sup_{\mathbf{X} \in \mathcal{S}_0} \{\underline{H}(\mathbf{Z})\} - \overline{H}(\mathbf{N}) \quad (92)$$

Next we upperbound the capacity as,
$$C(\mathbf{W}, \mathbf{g}, \mathbf{A}) \stackrel{(21)}{\leq} \sup_{\mathbf{X} \in \mathcal{S}_0} \{\underline{H}(\mathbf{Z}) - \underline{H}(\mathbf{Z}|\mathbf{X})\} \quad (93)$$
$$\stackrel{(89)}{=} \sup_{\mathbf{X} \in \mathcal{S}_0} \{\underline{H}(\mathbf{Z})\} - \underline{H}(\mathbf{N}) \quad (94)$$

By assumption $\underline{H}(\mathbf{N}) = \overline{H}(\mathbf{N})$ and combining (92) and (94) we have the desired result. ∎

### B. AWGN Example

The general formula of the previous section is now applied to the commonly found additive white Gaussian noise channel. The detector is motivated by the use of spread spectrum steganography[12], or more generally stochastic modulation[13].

The encoder-noise and attack-channel to be considered are additive white Gaussian noise (AWGN). For a stego-signal, $\mathbf{x} = (x_1, \ldots, x_n)$, the corrupted stego-signal is given by,
$$\mathbf{y} = (x_1 + n_1, \ldots, x_n + n_n),$$
where each $n_i \sim \mathcal{N}(0, \sigma_e^2)$, and all are independent.

The transition probabilities of the encoder-noise are given by,
$$W^n(\mathbf{y}|\mathbf{x}) = \frac{1}{(2\pi\sigma_e^2)^{\frac{n}{2}}} \exp\left\{-\frac{1}{2\sigma_e^2} \sum_{i=1}^{n}(y_i - x_i)^2\right\}. \quad (95)$$

Similarly, the attack-channel is AWGN as $\mathcal{N}(0, \sigma_a^2)$ so the transition probabilities are,
$$A^n(\mathbf{z}|\mathbf{y}) = \frac{1}{(2\pi\sigma_a^2)^{\frac{n}{2}}} \exp\left\{-\frac{1}{2\sigma_a^2} \sum_{i=1}^{n}(z_i - y_i)^2\right\}. \quad (96)$$

*1) Variance Steganalyzer:* In stochastic modulation, a pseudo-noise is modulated by a message and added to the cover-signal. This is done as the presence of noise in signal processing applications is a common occurrence.

If the passive adversary has knowledge of the distribution of the cover-signal and suspects stochastic modulation, they would expect the variance of a stego-signal will differ from a cover-signal. If the passive adversary knows the variance of the cover-distribution, they could design a steganalyzer to trigger if the variance of a test signal is higher than expected.

For example when testing the signal $\mathbf{y} = (y_1, \ldots, y_n)$ the variance steganalyzer operates as,

$$g_n(\mathbf{y}) = \begin{cases} 1, & \text{if } \frac{1}{n}\sum_{i=1}^{n} y_i^2 > c \\ 0, & \text{else} \end{cases} \quad (97)$$

That is to say, if the empirical variance of a test signal is above a certain threshold, the signal is considered steganographic.

*2) Additive Gaussian Channel Active Adversary:* In this section we derive the capacity under an active adversary. Assume that the adversary uses an additive i.i.d. Gaussian noise with variance $\sigma_a^2$ while the encoder noise is additive i.i.d. Gaussian with $\sigma_e^2$.

Let $\mathbf{N}_e = \{N_e\}^2$ where $N_e \sim \mathcal{N}(0, \sigma_e^2)$ and $\mathbf{N}_a = \{N_a\}$ where $N_a \sim \mathcal{N}(0, \sigma_a^2)$.

Let $\mathbf{N} = \mathbf{N}_e + \mathbf{N}_a = \{N^n = N_e^n + N_a^n\}_{n=1}^{\infty}$. Since both $\mathbf{N}_e$ and $\mathbf{N}_a$ are i.i.d. as $\mathcal{N}(0, \sigma_e^2)$ and $\mathcal{N}(0, \sigma_a^2)$, respectively, their sum is i.i.d. as $\mathcal{N}(0, \sigma_e^2 + \sigma_a^2)$, i.e. $\mathbf{N} = \{N\}$ with $N \sim \mathcal{N}(0, \sigma_e^2 + \sigma_a^2)$.

Since $\mathbf{N} = \{N\}$ with $N \sim \mathcal{N}(0, \sigma_e^2 + \sigma_a^2)$ we have the following relations,

$$\underline{H}(\mathbf{N}) = \overline{H}(\mathbf{N}) = H(N) = \frac{1}{2}\log 2\pi e\left(\sigma_a^2 + \sigma_e^2\right). \quad (98)$$

Since $\underline{H}(\mathbf{N}) = \overline{H}(\mathbf{N})$ we see that the noise sequence satisfies the strong converse property.

*3) Active Adversary Capacity:* We now derive the secure capacity of the above stego-channel. Since the noises are i.i.d., the general sequence $\mathbf{N}$ will satisfy the strong converse and allow the use of Theorem 4.1.

The formal proof is then followed by a discussion of the results and a description using the classic sphere packing intuition.

*Theorem 4.2:* For the stego-channel $(\mathbf{W}, \mathbf{g}, \mathbf{A}) = \{(W^n, g_n, A^n)\}_{n=1}^{\infty}$ with $W^n$ and $A^n$ defined by (95) and (96) respectively, and $g_n$ defined by (97) the secure capacity is,

$$C(\mathbf{W}, \mathbf{g}, \mathbf{A}) = \frac{1}{2}\log\frac{c + \sigma_a^2}{\sigma_e^2 + \sigma_a^2}. \quad (99)$$

*Proof:* From Theorem 4.1 and (98) we have,

$$C(\mathbf{W}, \mathbf{g}, \mathbf{A}) = \sup_{\mathbf{X} \in \mathcal{S}_0} \{\underline{H}(\mathbf{Z})\} - \overline{H}(\mathbf{N}) \quad (100)$$

$$= \sup_{\mathbf{X} \in \mathcal{S}_0} \{\underline{H}(\mathbf{Z})\} - \frac{1}{2}\log 2\pi e\left(\sigma_a^2 + \sigma_e^2\right). \quad (101)$$

**Achievability:**
Let $\overline{\mathbf{X}} = \{\overline{X}\}$ where $\overline{X} \sim \mathcal{N}(0, c - \sigma_e^2)$. Thus $\overline{\mathbf{Y}} = \overline{\mathbf{X}} + \mathbf{N}_e = \{\overline{Y}\}$ with $\overline{Y} = \overline{X} + N_e$. By addition of independent Gaussians, $\overline{Y} \sim \mathcal{N}(0, c)$. This gives,

$$\Pr\left\{\frac{1}{n}\sum_{i=1}^{n}\left(\overline{Y}_i^{(n)}\right)^2 > c\right\} \to 0, \quad (102)$$

---

[2]Recall that for a general sequence, $\mathbf{X} = \{X^n = (X_1^{(n)}, \ldots, X_n^{(n)})\}_{n=1}^{\infty}$ when $\mathbf{X} = \{X\}$ is written it means that each $X_i^{(n)}$ is independent and identically distributed as $X$.

and we see that $\overline{\mathbf{X}} \in \mathcal{S}_0$. Similarly, $\overline{\mathbf{Z}} = \mathbf{N}_a + \overline{\mathbf{Y}} = \{\overline{Z}\}$ with $\overline{Z} = \overline{X} + N_e + N_a$. Again by addition of independent Gaussians we have $\overline{Z} \sim \mathcal{N}(0, c + \sigma_a^2)$.

This allows for a lower bound of,

$$C(\mathbf{W}, \mathbf{g}, \mathbf{A}) \stackrel{(101)}{=} \sup_{\mathbf{X} \in \mathcal{S}_0} \underline{H}(\mathbf{Z}) - \frac{1}{2}\log\left(2\pi e(\sigma_e^2 + \sigma_a^2)\right) \quad (103a)$$

$$\geq \underline{H}(\overline{\mathbf{Z}}) - \frac{1}{2}\log\left(2\pi e(\sigma_e^2 + \sigma_a^2)\right) \quad (103b)$$

$$= \frac{1}{2}\log\frac{c + \sigma_a^2}{\sigma_e^2 + \sigma_a^2} \quad (103c)$$

**Converse:**
To find the upperbound we will make use of a number of simple lemmas:

*Lemma 4.1:* For a given stego-channel with secure input distribution set $\mathcal{S}_0$ and secure output distribution set $\mathcal{T}_0$, the following holds,

$$\sup_{\mathbf{X} \in \mathcal{S}_0} \underline{H}(\mathbf{Z}) \leq \sup_{\mathbf{Y} \in \mathcal{T}_0} \underline{H}(\mathbf{Z}). \quad (104)$$

*Proof:* By definition for any $\mathbf{X} \in \mathcal{S}_0$ and $\mathbf{X} \xrightarrow{\mathbf{W}} \mathbf{Y}$, we have $\mathbf{Y} \in \mathcal{T}_0$. ∎

*Lemma 4.2:* For $Y^n = (Y_1^{(n)}, Y_2^{(n)}, \ldots, Y_n^{(n)})$ let $K_{ij}^{(n)}$ be the covariance between $Y_i^{(n)}$ and $Y_j^{(n)}$, that is $K_{ij}^{(n)} := E\left\{Y_i^{(n)}Y_j^{(n)}\right\}$. For the stego-channel defined above, if $\mathbf{Y} = \{Y^n\}_{n=1}^{\infty} \in \mathcal{T}_0$ we have for any $\gamma > 0$ there exists some $N$ such that for all $n > N$,

$$\frac{1}{n}\sum_{i=1}^{n} K_{ii}^{(n)} + \sigma_a^2 < c + \sigma_a^2 + \gamma. \quad (105)$$

*Proof:* It suffices to show,

$$\frac{1}{n}\sum_{i=1}^{n} K_{ii}^{(n)} < c + \gamma, \quad (106)$$

for all $n$ greater than some $N$.

To show this, assume that no such $N$ exists, thus we have a subsequence $n_k$ such that,

$$\frac{1}{n_k}\sum_{i=1}^{n_k} K_{ii}^{(n_k)} \geq c + \gamma. \quad (107)$$

This means that,

$$\frac{1}{n_k}\sum_{i=1}^{n_k} K_{ii}^{(n_k)} = E\left\{\frac{1}{n_k}\sum_{i=1}^{n_k} y_i^2\right\} \geq c + \gamma,$$

which in turn implies that,

$$\Pr\left\{g_{n_k}(Y^{n_k}) = 0\right\} \to 0.$$

This is a contradiction as it shows $\mathbf{Y} = \{Y^n\}_{n=1}^{\infty} \notin \mathcal{T}_0$. ∎

*Lemma 4.3:* For any $Z^n = (Z_1, \ldots, Z_n)$ with $C_{ij} = E\{Z_iZ_j\}$,

$$H(Z^n) \leq \frac{1}{2}\log(2\pi e)^n\left(\frac{1}{n}\sum_{i=1}^{n} C_{ii}\right)^n. \quad (108)$$

*Proof:* From [14, Chap. 9.6] we have,

$$H(Z^n) \leq \frac{1}{2}\log(2\pi e)^n \prod_{i=1}^{n} C_{ii}. \tag{109}$$

The result follows from application of the arithmetic-geometric inequality. ∎

*Lemma 4.4:* For the above stego-channel, any $\mathbf{Y} \in \mathcal{T}_0$ and any $\epsilon > 0$ we have,

$$\liminf_{n \to \infty} \frac{1}{n} H(Z^n) < \frac{1}{2}\log 2\pi e(c + \sigma_a^2) + \epsilon, \tag{110}$$

where $\mathbf{Z} = \{Z^n\}_{n=1}^{\infty}$ and $\mathbf{Y} \xrightarrow{\mathbf{A}} \mathbf{Z}$.

*Proof:* Let any $\epsilon > 0$ be given and choose $\gamma > 0$ such that,

$$\gamma \leq (c + \sigma_a^2)(e^{2\epsilon} - 1),$$

this gives,

$$\frac{1}{2}\log 2\pi e(c + \sigma_a^2 + \gamma) \leq \frac{1}{2}\log 2\pi e(c + \sigma_a^2) + \epsilon. \tag{111}$$

Letting $C_{ij}^{(n)} = E\left\{Z_i^{(n)} Z_j^{(n)}\right\}$ and $K_{ij}^{(n)} = E\left\{Y_i^{(n)} Y_j^{(n)}\right\}$ we note that $Z_i^{(n)} = Y_i^{(n)} + N_a$. This gives,

$$C_{ii}^{(n)} = K_{ii}^{(n)} + \sigma_a^2. \tag{112}$$

This gives,

$$\frac{1}{n}H(Z^n) \stackrel{(L4.3)}{\leq} \frac{1}{2n}\log(2\pi e)^n \left(\frac{1}{n}\sum_{i=1}^{n} C_{ii}^{(n)}\right)^n \tag{113}$$

$$\stackrel{(112)}{=} \frac{1}{2n}\log(2\pi e)^n \left(\frac{1}{n}\sum_{i=1}^{n} K_{ii}^{(n)} + \sigma_a^2\right)^n \tag{114}$$

$$\stackrel{(L4.2)}{<} \frac{1}{2n}\log(2\pi e)^n (c + \sigma_a^2 + \gamma)^n \tag{115}$$

$$\stackrel{(111)}{\leq} \frac{1}{2}\log 2\pi e(c + \sigma_a^2) + \epsilon \tag{116}$$

The inequality of (115) holds for all but a finite number of $n$ by Lemma 4.2. ∎

We now show the upperbound:

Beginning with the specialization of Theorem 4.1,

$$C(\mathbf{W}, \mathbf{g}, \mathbf{A}) \stackrel{(101)}{=} \sup_{\mathbf{X} \in \mathcal{S}_0} \{\underline{H}(\mathbf{Z})\} - \frac{1}{2}\log 2\pi e(\sigma_e^2 + \sigma_a^2) \tag{117a}$$

$$\stackrel{(L4.1)}{\leq} \sup_{\mathbf{Y} \in \mathcal{T}_0} \{\underline{H}(\mathbf{Z})\} - \frac{1}{2}\log 2\pi e(\sigma_e^2 + \sigma_a^2) \tag{117b}$$

$$\stackrel{(20)}{\leq} \sup_{\mathbf{Y} \in \mathcal{T}_0} \liminf_{n \to \infty} \frac{1}{n} H(Z^n)$$
$$- \frac{1}{2}\log 2\pi e(\sigma_e^2 + \sigma_a^2) \tag{117c}$$

$$\stackrel{(L4.4)}{<} \frac{1}{2}\log \frac{c + \sigma_a^2}{\sigma_e^2 + \sigma_a^2} + \epsilon \tag{117d}$$

Combining (103c) and (117d) we have for any $\epsilon > 0$,

$$\frac{1}{2}\log \frac{c + \sigma_a^2}{\sigma_e^2 + \sigma_a^2} \leq C(\mathbf{W}, \mathbf{g}, \mathbf{A}) < \frac{1}{2}\log \frac{c + \sigma_a^2}{\sigma_e^2 + \sigma_a^2} + \epsilon,$$

and we see that $C(\mathbf{W}, \mathbf{g}, \mathbf{A}) = \frac{1}{2}\log \frac{c + \sigma_a^2}{\sigma_e^2 + \sigma_a^2}$. ∎

TABLE III
GAUSSIAN ADDITIVE NOISE CAPACITIES

| Channel | Secure Capacity | Encoder Noise | Attack Noise |
|---|---|---|---|
| $C(\mathbf{W}, \mathbf{g}, \mathbf{A})$ | $\frac{1}{2}\log \frac{c+\sigma_a^2}{\sigma_e^2+\sigma_a^2}$ | $\sigma_e^2$ | $\sigma_a^2$ |
| $C(\mathbf{W}, \mathbf{g})$ | $\frac{1}{2}\log \frac{c}{\sigma_e^2}$ | $\sigma_e^2$ | 0 |
| $C(\cdot, \mathbf{g}, \mathbf{A})$ | $\frac{1}{2}\log \frac{c+\sigma_a^2}{\sigma_a^2}$ | 0 | $\sigma_a^2$ |
| $C(\cdot, \mathbf{g})$ | $\lim_{\sigma^2 \to 0} \frac{1}{2}\log \frac{c+\sigma^2}{2\sigma^2}$ | 0 | 0 |

*4) Noise Cases:* We now use this theorem to investigate the behavior of the capacity under different noise conditions.

*5) Large Attack Case:* We first consider the case where $\sigma_a^2$ is much larger than both $c$ and $\sigma_e^2$. This gives,

$$C(\mathbf{W}, \mathbf{g}, \mathbf{A}) = \frac{1}{2}\log \frac{c + \sigma_a^2}{\sigma_e^2 + \sigma_a^2} \approx \frac{1}{2}\log \frac{\sigma_a^2}{\sigma_a^2} = 0.$$

This shows that when the attack noise is large enough, the capacity of the stego-channel goes to zero. Intuitively this is due to the fact that the variance steganalyzer places a power constraint (of $c$) on any signals it allows to pass. If the attack noise is much larger than $c$, a message simply cannot be transmitted with enough power to overcome that noise and $\epsilon_n \to 0$ is impossible.

*6) Large Encoder-Noise Case:* Next we consider the case where $\sigma_e^2 \geq c$.

Since $\frac{c+\sigma_a^2}{\sigma_e^2+\sigma_a^2} \leq 1$, we have $\log \frac{c+\sigma_a^2}{\sigma_e^2+\sigma_a^2} \leq 0$. This gives,

$$C(\mathbf{W}, \mathbf{g}, \mathbf{A}) = \frac{1}{2}\log \frac{c + \sigma_a^2}{\sigma_e^2 + \sigma_a^2} \leq 0$$

As capacity is always greater or equal to zero, we see that the capacity of this system is indeed zero. This is because no matter what codeword is sent, the encoder-noise will corrupt it into the impermissible set and the steganalyzer will be triggered, that is $\delta_n \to 0$ is impossible.

This case illuminates the importance of the additional constraint in communication over a stego-channel, as even if $\epsilon \to 0$ the capacity of the stego-channel is still zero.

*7) Noiseless Case:* Consider the noiseless case where $\sigma_e^2 = \sigma_a^2 = \sigma^2$ and $\sigma^2 \to 0$. This gives,

$$\lim_{\sigma^2 \to 0} C(\mathbf{W}, \mathbf{g}, \mathbf{A}) = \lim_{\sigma^2 \to 0} \frac{1}{2}\log \frac{c + \sigma^2}{\sigma^2 + \sigma^2} = \infty.$$

Since the channel is noiseless and the permissible set size is infinite (as well as input and output alphabets), the capacity is unbounded.

*8) Geometric Intuition:* In this section we present some geometric intuition to the previous results, similar to the case of the classic additive Gaussian noise[14], [15].

We will consider the case of only an encoder-noise of $\sigma^2$, shown in Figure 9.

From the above theorem we see that,

$$C(\mathbf{W}, \mathbf{g}) = \frac{1}{2}\log \frac{c}{\sigma^2}. \tag{118}$$

The most basic element will be the volume of an $n$ dimensional sphere of radius $r$. In this case, the volume is equal to $A_n r^n$ where $A_n$ is a constant dependent only on the dimension $n$.

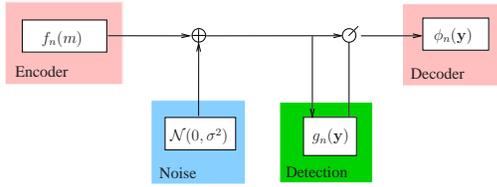

Fig. 9. AWGN Channel Passive Adversary

The fundamental question is: what is the capacity of the stego-channel, or how many codewords can we reliably and safely use? To answer this, we must consider the two constraints on a secure system: *error probability* and *detection probability*.

*9) Error Probability:* Since we have that $\mathcal{X}^n = \mathcal{Y}^n = \Re^n$, we may view each codeword as a point in $\Re^n$. When we transmit a given codeword, we may think of the addition of noise as moving the point around in that space. As the power of the noise is $\sigma^2$, the probability that the received codeword has moved more than $\sqrt{n\sigma^2}$ away from where it started goes to zero as $n \to \infty$. This means a received codeword will likely be contained in a sphere of radius $\sqrt{n\sigma^2}$ centered on the transmitted codeword. If we receive a signal inside such a sphere, it is likely that the transmitted codeword was the center of that sphere. In this manner we can define a coding system by choosing the codewords such that their spheres do not overlap. This results in no confusion during decoding and achieves the requirement of vanishing error probability.

*10) Detection Probability:* We begin by looking at the permissible set. The permissible set for our $g_n$ is given by,

$$\mathcal{P}_{g_n} = \{\mathbf{y} \in \mathcal{Y}^n : \sum_{i=1}^n y_i^2 < nc\}. \quad (119)$$

Clearly the permissible set is a sphere of radius $\sqrt{nc}$ centered at the origin. If a test signal falls inside this sphere it is classified as non-steganographic, whereas if it is outside it is considered steganographic.

The second criteria for a secure system is that the probability of detection goes to zero. If we were to place each codeword such that its sphere was inside the permissible set, we know that the probability of detection will go to zero.

*11) Capacity:* From the above, we know that the codeword spheres cannot overlap (to ensure no errors). We also know that all the codeword spheres must fit inside the permissible set (to ensure no detection). If we calculate the number of non-overlapping spheres we may pack into the permissible set, we will have a general idea of the number of codewords we can use.

Since the volume of the permissible set is $A_n(nc)^{\frac{n}{2}}$ and the volume of each codeword sphere is $A_n(n\sigma^2)^{\frac{n}{2}}$, we can place approximately,

$$\frac{A_n(nc)^{\frac{n}{2}}}{A_n(n\sigma^2)^{\frac{n}{2}}} = \left(\frac{c}{\sigma^2}\right)^{\frac{n}{2}},$$

non-overlapping sphere inside the permissible set.

Using the center of each sphere as a codeword, we have $M_n$ codewords where $M_n = \left(\frac{c}{\sigma^2}\right)^{\frac{n}{2}}$.

If we consider the capacity as $C(\mathbf{W}, \mathbf{g}) = \lim \frac{1}{n} \log M_n$ we have,

$$C(\mathbf{W}, \mathbf{g}) = \lim \frac{1}{n} \log \left(\frac{c}{\sigma^2}\right)^{\frac{n}{2}} \quad (120a)$$
$$= \frac{1}{2} \log \frac{c}{\sigma^2}, \quad (120b)$$

which agrees with the result of Theorem 4.2.

## V. Previous Work Revisited

### A. Cachin Perfect Security

In Cachin's definition of perfect security[16], the cover-signal distribution and the stego-signal distribution are each required to be independent and identically distributed. This gives the following secure-input set,

$$\mathcal{S}_0 = \left\{\mathbf{X} = \{X\} : \lim_{n\to\infty} \frac{1}{n} D(S^n || X^n) = 0\right\}. \quad (121)$$

The i.i.d. property means that $D(S^n||X^n) = nD(S||X)$ so we see that the above is equivalent to,

$$\mathcal{S}_0 = \{\mathbf{X} = \{X\} : D(S||X) = 0\} \quad (122)$$
$$= \{\mathbf{X} = \{X\} : p_S = p_X\} \quad (123)$$

Since Cachin's definition does not model noise, we may consider it as noiseless and apply Theorem 3.1,

$$C(\mathbf{W}, \mathbf{g}) = \sup_{\mathbf{X} \in \mathcal{S}_0} \underline{H}(\mathbf{X}) = H(S). \quad (124)$$

This result states that in a system that is perfectly secure (in Cachin's definition), the limit on the amount of information that may be transferred each channel use is equal to the entropy of the source. This is intuitive because in Cachin's definition the output distribution of the encoder is constrained to be equal to the cover-signal distribution.

### B. Empirical Distribution Steganalyzer

The *empirical distribution steganalyzer* is motivated by the fact that the empirical distribution from a stationary memoryless source converges to the actual distribution of that source. Accordingly, if the empirical distribution of the test signal converges to the cover-signal distribution it is considered to be non-steganographic.

Assume that $p_S$ is a discrete distribution over the finite alphabet $\mathcal{S}$. Let a sequence, $\{s^n\}_{n=1}^\infty$ with each $s^n \in \mathcal{S}^n$ be used to specify the steganalyzer for a test signal $\mathbf{x}$ as,

$$g_n(\mathbf{x}) = \begin{cases} 0 & \text{if } P_{[s^n]} = P_{[\mathbf{x}]}, \\ 1 & \text{if } P_{[s^n]} \neq P_{[\mathbf{x}]}. \end{cases} \quad (125)$$

where $P_{[\mathbf{x}]}$ is the empirical distribution of $\mathbf{x}$.

The permissible set for $g_n$ is equal to the type class of $P_{[s^n]}$, i.e.,

$$\mathcal{P}_{g_n} = T(P_{[s^n]}) := \{\mathbf{x} \in \mathcal{X}^n : P_{[\mathbf{x}]} = P_{[s^n]}\}. \quad (126)$$

*Theorem 5.1 (Empircal Distribution Steganalyzer Capacity):*

$$C(\mathbf{W}, \mathbf{g}) = H(S). \quad (127)$$

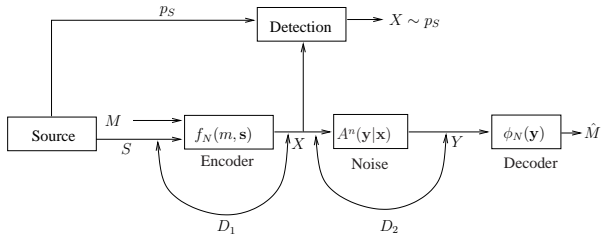

Fig. 10. Moulin Stego-channel

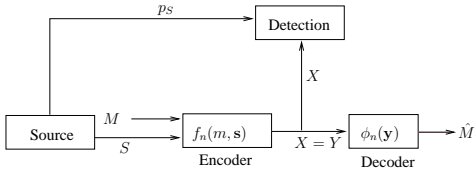

Fig. 11. Equivalent Stego-channel

*Proof:* Since the channel is noiseless we may apply Theorem 3.1.

$$C(\mathbf{W}, \mathbf{g}) = \liminf_{n \to \infty} \frac{1}{n} \log |\mathcal{P}_{g_n}| \qquad (128a)$$
$$= \liminf_{n \to \infty} \frac{1}{n} \log |T(s^n)| \qquad (128b)$$
$$= H(S) \qquad (128c)$$

Here we have used the fact that the permissible set for the empirical distribution detection function is the type class in (128b). Additionally, by Varadarjan's Theorem[17], $P_{[s^n]}(x) \to p_S(x)$ almost surely (here the convergence is uniform in $x$ as well). This allows for the use of the type class-entropy bound from Theorem D.1 that provides the final result. ∎

### C. Moulin Steganographic Capacity

Moulin's formulation[2], [3] of the stego-channel is shown in Figure 10. This is somewhat different than the formulation shown in Figure 1; most notable is the presence of distortion constraints and an absence of a distortion function prior to the steganalyzer. Additionally, an explicit steganalyzer is not defined and a hypothetical $\mathbf{X} \sim p_S$ is used. In order to have the two formulations coincide a number of simplifications are needed for each model.

For our model,
- The stego-channel is noiseless
- The steganalyzer is the empirical distribution

For Moulin's model,
- Passive adversary ($D_2 = 0$)
- No distortion constraint on encoder ($D_1 = \infty$)

These changes produce the stego-channel shown in Figure 11.

*Theorem 5.2:* For the stego-channel shown in Figure 11, the capacities of this work and Moulin's agree,

$$C(\mathbf{W}, \mathbf{g}) = C^{STEG}(\infty, 0) = H(S). \qquad (129)$$

*Proof:* Theorem 5.1 shows $C(\mathbf{W}, \mathbf{g}) = H(S)$.

We now show Moulin's capacity is equal to this value. In the case of a passive adversary ($D_2 = 0$), the following is the capacity of the stego-channel[2],

$$C^{STEG}(D_1, 0) = \sup_{Q' \in \mathcal{Q}'} H(X|S) \qquad (130)$$

where a $p \in \mathcal{Q}'$ is feasible if,

$$\sum_{s,x} p(x|s) p_S(s) d(s,x) \leq D_1, \qquad (131)$$

and

$$\sum_s p(x|s) p_S(s) = p_S(x). \qquad (132)$$

The capacity can be found for unbounded $D_1$ as,

$$C^{STEG}(\infty, 0) = \sup_{p(x|s) \in \mathcal{Q}'} H(X|S) \qquad (133a)$$
$$= H(S) - \min_{p(x|s) \in \mathcal{Q}'} I(S; X) \qquad (133b)$$
$$= H(S) \qquad (133c)$$

where the final line comes from choosing $p(x) = p_S(x)$. ∎

## VI. CONCLUSIONS

A framework for evaluating the capacity of steganographic channels under an active adversary has been introduced. The system considers a noise corrupting the signal before the detection function in order to model real-world distortions such as compression, quantization, etc.

Constraints on the encoder dealing with distortion and a cover-signal are not considered. Instead, the focus is to develop the theory necessary to analyze the interplay between the channel and detection function that results in the steganographic capacity.

The method uses an information-spectrum approach that allows for the analysis of arbitrary detection functions and channels. This provides machinery necessary to analyze a very broad range of steganographic channels.

In addition to offering insight into the limits of performance for steganographic algorithms, this formulation of capacity can be used to analyze a different and fundamentally important facet of steganalysis. While false alarms and missed signals have rightfully dominated the steganalysis literature, very little is known about the amount of information that can be sent past these algorithms. This work presents a theory to shed light onto this important quantity called steganographic capacity.

## APPENDIX A
### $\epsilon$-STRONG CONVERSE PROOF

A stego-channel $(\mathbf{W}, \mathbf{g}, \mathbf{A})$ satisfies the $\epsilon$-strong converse property (for a fixed $\delta$) if and only if,

$$\sup_{\mathbf{X} \in \mathcal{S}_\delta} \underline{I}(\mathbf{X}; \mathbf{Z}) = \sup_{\mathbf{X} \in \mathcal{S}_\delta} \overline{I}(\mathbf{X}; \mathbf{Z}). \qquad (A.134)$$

*Proof:* First assume $\sup_{\mathbf{X} \in \mathcal{S}_\delta} \underline{I}(\mathbf{X}; \mathbf{Z}) = \sup_{\mathbf{X} \in \mathcal{S}_\delta} \overline{I}(\mathbf{X}; \mathbf{Z})$. Let $R = C(0, \delta | \mathbf{W}, \mathbf{g}, \mathbf{A}) + 3\gamma$ with $\gamma > 0$. Consider an $(n, M_n, \epsilon_n, \delta_n)$-code with,

$$\liminf_{n \to \infty} \frac{1}{n} \log M_n \geq R,$$

and
$$\limsup_{n\to\infty} \delta_n \leq \delta.$$

Let $\mathbf{X}$ represent the uniform input due to this code and $\mathbf{Z}$ the output after the channel $\mathbf{Q} = \mathbf{AX}$. From the Feinstein Dual [6], [7] we know,
$$\epsilon_n \geq \Pr\left\{\frac{1}{n}i(X^n; Z^n) \leq \frac{1}{n}\log M_n - \gamma\right\} - e^{-n\gamma}. \quad (A.135)$$

We also know there exists $n_0$ such that for all $n > n_0$ that,
$$\frac{1}{n}\log M_n \geq R - \gamma, \quad (A.136)$$

so for $n > n_0$,
$$\epsilon_n \geq \Pr\left\{\frac{1}{n}i(X^n; Z^n) \leq R - 2\gamma\right\} - e^{-n\gamma}. \quad (A.137)$$

We now show that the probability term above tends to 1.

Using Theorem 2.2 we have,
$$R = C(0, \delta|\mathbf{W}, \mathbf{g}, \mathbf{A}) + 3\gamma \quad (A.138)$$
$$= \sup_{\mathbf{X}\in\mathcal{S}_\delta} \underline{I}(\mathbf{X}; \mathbf{Z}) + 3\gamma \quad (A.139)$$
$$= \sup_{\mathbf{X}\in\mathcal{S}_\delta} \overline{I}(\mathbf{X}; \mathbf{Z}) + 3\gamma \quad (A.140)$$

Rewriting gives,
$$R - 2\gamma = \sup_{\mathbf{X}\in\mathcal{S}_\delta} \overline{I}(\mathbf{X}; \mathbf{Z}) + \gamma. \quad (A.141)$$

By the definition of $\overline{I}(\mathbf{X}; \mathbf{Z})$ we finally have,
$$\lim_{n\to\infty} \Pr\left\{\frac{1}{n}i(X^n; Z^n) \leq R - 2\gamma\right\} = 1, \quad (A.142)$$

which together with A.137 shows that that $\lim_{n\to\infty} \epsilon_n = 1$.

For the other direction assume,
$$\lim_{n\to\infty} \epsilon_n = 1, \quad (A.143)$$

and,
$$\limsup_{n\to\infty} \delta_n \leq \delta. \quad (A.144)$$

Set $R = C(0, \delta|\mathbf{W}, \mathbf{g}, \mathbf{A}) + \gamma$ for any $\gamma > 0$ and set $M_n = e^{nR}$. Clearly,
$$\liminf_{n\to\infty} \frac{1}{n}\log M_n = R > C(0, \delta|\mathbf{W}, \mathbf{g}, \mathbf{A}).$$

For any $\mathbf{X} \in \mathcal{S}_\delta$ (and its corresponding $\mathbf{Z}$), using Feinstein's Lemma [11] we have an $(n, M_n, \epsilon_n)$-code satisfying,
$$\epsilon_n \leq \Pr\left\{\frac{1}{n}i(X^n; Z^n) \leq R + \gamma\right\} + e^{-n\gamma}. \quad (A.145)$$

From the error assumption we see that,
$$\lim_{n\to\infty} \Pr\left\{\frac{1}{n}i(X^n; Z^n) \leq R + \gamma\right\} = 1. \quad (A.146)$$

This means that,
$$R + \gamma \geq \overline{I}(\mathbf{X}; \mathbf{Z}), \quad (A.147)$$

and since $\mathbf{X} \in \mathcal{S}_\delta$ is arbitrary we have,
$$R + \gamma \geq \sup_{\mathbf{X}\in\mathcal{S}_\delta} \overline{I}(\mathbf{X}; \mathbf{Z}). \quad (A.148)$$

Substituting we have that,
$$\sup_{\mathbf{X}\in\mathcal{S}_\delta} \overline{I}(\mathbf{X}; \mathbf{Z}) \leq R + \gamma \quad (A.149)$$
$$= C(0, \delta|\mathbf{W}, \mathbf{g}, \mathbf{A}) + 2\gamma \quad (A.150)$$
$$= \sup_{\mathbf{X}\in\mathcal{S}_\delta} \underline{I}(\mathbf{X}; \mathbf{Z}) + 2\gamma \quad (A.151)$$

As $\gamma$ is arbitrarily close to 0 we have,
$$\sup_{\mathbf{X}\in\mathcal{S}_\delta} \overline{I}(\mathbf{X}; \mathbf{Z}) \leq \sup_{\mathbf{X}\in\mathcal{S}_\delta} \underline{I}(\mathbf{X}; \mathbf{Z}). \quad (A.152)$$

Also, by definition,
$$\sup_{\mathbf{X}\in\mathcal{S}_\delta} \overline{I}(\mathbf{X}; \mathbf{Z}) \geq \sup_{\mathbf{X}\in\mathcal{S}_\delta} \underline{I}(\mathbf{X}; \mathbf{Z}), \quad (A.153)$$

showing equality and completing the proof. ∎

## APPENDIX B
### SPECTRAL INF-ENTROPY BOUND

For a discrete $\mathbf{g} = \{\mathcal{P}_n\}_{n=1}^\infty$ with corresponding secure output set $\mathcal{T}_0$,
$$\sup_{\mathbf{Y}\in\mathcal{T}_0} \underline{H}(\mathbf{Y}) = \liminf_{n\to\infty} \frac{1}{n}\log|\mathcal{P}_n|$$

*Proof:* Let $\mathcal{U}(A)$ represent the uniform distribution on a set $A$.

Since $\mathbf{Y}^* = \{\mathcal{U}(\mathcal{P}_n)\}_{i=1}^\infty \in \mathcal{T}_0$ we have,
$$\sup_{\mathbf{Y}\in\mathcal{T}_0} \underline{H}(\mathbf{Y}) \geq \underline{H}(\mathbf{Y}^*) = \liminf_{n\to\infty} \frac{1}{n}\log|\mathcal{P}_n| \quad (B.154)$$

Now assume there exists $\overline{\mathbf{Y}} \in \mathcal{T}_0$ with $\overline{\mathbf{Y}} = \{\bar{Y}^n\}_{n=1}^\infty$, such that,
$$\underline{H}(\overline{\mathbf{Y}}) = \underline{H}(\mathbf{Y}^*) + 3\gamma, \quad (B.155)$$

for any $\gamma > 0$.

This means that,
$$\lim_{n\to\infty} \Pr\left\{\frac{1}{n}\log\frac{1}{p_{\bar{Y}^n}(\bar{Y}^n)} < \underline{H}(\mathbf{Y}^*) + 2\gamma\right\} = 0 \quad (B.156)$$

By (B.154) we have $\underline{H}(\mathbf{Y}^*) = \liminf_{n\to\infty} \frac{1}{n}\log|\mathcal{P}_n|$ and from the definition of lim inf we may find a subsequence indexed by $k_n$ such that,
$$\underline{H}(\mathbf{Y}^*) + 2\gamma \geq \frac{1}{k_n}\log|\mathcal{P}_{k_n}| + \gamma. \quad (B.157)$$

For any $k_n$ (B.157) holds and we have,
$$\Pr\left\{\frac{1}{k_n}\log\frac{1}{p_{\bar{Y}^{k_n}}(\bar{Y}^{k_n})} < \frac{1}{k_n}\log|\mathcal{P}_{k_n}| + \gamma\right\} \leq$$
$$\Pr\left\{\frac{1}{k_n}\log\frac{1}{p_{\bar{Y}^{k_n}}(\bar{Y}^{k_n})} < \underline{H}(\mathbf{Y}^*) + 2\gamma\right\}. \quad (B.158)$$

Applying this result to (B.156) we have,
$$\lim_{n\to\infty} \Pr\left\{\frac{1}{k_n}\log\frac{1}{p_{\bar{Y}^{k_n}}(\bar{Y}^{k_n})} < \frac{1}{k_n}\log|\mathcal{P}_{k_n}| + \gamma\right\} = 0. \quad (B.159)$$

For any $\epsilon > 0$ and $n$ greater than some $n_0$,
$$\Pr\left\{p_{\bar{Y}^{k_n}}(\bar{Y}^{k_n}) > \frac{e^{-k_n\gamma}}{\mathcal{P}_{k_n}}\right\} < \epsilon. \quad (B.160)$$

Let,

$$A_{k_n} = \left\{ \mathbf{y} \in \mathcal{Y}^n : p_{\bar{Y}^{k_n}}(\bar{Y}^{k_n}) > \frac{e^{-k_n\gamma}}{|\mathcal{P}_{k_n}|} \right\}, \quad \text{(B.161)}$$

and for all $n > n_0$, we have $p_{\bar{Y}^{k_n}}(A_{k_n}) < \epsilon$.

For $n > n_0$ we may calculate the probability of the permissible set (for the subsequence) as,

$$p_{\bar{Y}^{k_n}}(\mathcal{P}_{k_n}) = \sum_{\mathbf{y} \in \mathcal{P}_{k_n}} p_{\bar{Y}^{k_n}}(\mathbf{y}) \quad \text{(B.162a)}$$

$$= \sum_{\mathbf{y} \in \mathcal{P}_{k_n} \cap A_{k_n}^c} p_{\bar{Y}^{k_n}}(\mathbf{y}) + \sum_{\mathbf{y} \in \mathcal{P}_{k_n} \cap A_{k_n}} p_{\bar{Y}^{k_n}}(\mathbf{y}) \quad \text{(B.162b)}$$

$$\leq \sum_{\mathbf{y} \in \mathcal{P}_{k_n}} \frac{e^{-k_n\gamma}}{|\mathcal{P}_{k_n}|} + \sum_{\mathbf{y} \in A_{k_n}} p_{\bar{Y}^{k_n}}(\mathbf{y}) \quad \text{(B.162c)}$$

$$< e^{-k_n\gamma} + \epsilon \quad \text{(B.162d)}$$

This shows $p_{\bar{Y}^{k_n}}(\mathcal{P}_{k_n}) \not\to 1$ and we have a contradiction as $\overline{\mathbf{Y}} \notin \mathcal{T}_0$. ∎

## APPENDIX C
## SPECTRAL SUP-ENTROPY BOUND

For discrete $\mathbf{g} = \{\mathcal{P}_n\}_{n=1}^\infty$ with corresponding secure output set $\mathcal{T}_0$,

$$\sup_{\mathbf{Y} \in \mathcal{T}_0} \overline{H}(\mathbf{Y}) = \limsup_{n \to \infty} \frac{1}{n} \log |\mathcal{P}_n|$$

*Proof:* Since $\mathbf{Y}^* = \{\mathcal{U}(\mathcal{P}_n)\}_{i=1}^\infty \in \mathcal{T}_0$ we have,

$$\sup_{\mathbf{Y} \in \mathcal{T}_0} \overline{H}(\mathbf{Y}) \geq \overline{H}(\mathbf{Y}^*) \quad \text{(C.163a)}$$

$$= \limsup_{n \to \infty} \frac{1}{n} \log |\mathcal{P}_n| \quad \text{(C.163b)}$$

Now assume there exists $\overline{\mathbf{Y}} \in \mathcal{T}_0$, with $\overline{\mathbf{Y}} = \{\bar{Y}^n\}_{n=1}^\infty$ such that,

$$\overline{H}(\overline{\mathbf{Y}}) = \overline{H}(\mathbf{Y}^*) + \frac{\gamma}{4}, \quad \text{(C.164)}$$

for any $\gamma > 0$.

This means that,

$$\lim_{n \to \infty} \Pr\left\{ \frac{1}{n} \log \frac{1}{p_{\bar{Y}^n}(\bar{Y}^n)} > \overline{H}(\mathbf{Y}^*) + \frac{\gamma}{2} \right\} = 0 \quad \text{(C.165)}$$

By the definition of $\limsup$ for some subsequence $k_n$ we have,

$$\frac{1}{k_n} \log |\mathcal{P}_{k_n}| + \gamma > \overline{H}(\mathbf{Y}^*) + \frac{\gamma}{2} \quad \text{(C.166)}$$

and

$$\lim_{n \to \infty} \Pr\left\{ \frac{1}{k_n} \log \frac{1}{p_{\bar{Y}^{k_n}}(\bar{Y}^{k_n})} > \frac{1}{k_n} \log |\mathcal{P}_{k_n}| + \gamma \right\} = 0. \quad \text{(C.167)}$$

For any $\epsilon > 0$ letting,

$$A_{k_n} = \left\{ \mathbf{y} \in \mathcal{X}^n : p_{\bar{Y}^{k_n}}(\bar{Y}^{k_n}) < \frac{e^{-k_n\gamma}}{|\mathcal{P}_{k_n}|} \right\} \quad \text{(C.168)}$$

we may find $n_0$ where for $n > n_0$,

$$p_{\bar{Y}^{k_n}}(A_{k_n}) < \epsilon. \quad \text{(C.169)}$$

For $n > n_0$ the probability of the permissible set (in this subsequence) is,

$$p_{\bar{Y}^{k_n}}(\mathcal{P}_{k_n}) = \sum_{\mathbf{x} \in \mathcal{P}_{k_n}} p_{\bar{Y}^{k_n}}(\mathbf{y}) \quad \text{(C.170a)}$$

$$= \sum_{\mathbf{y} \in \mathcal{P}_{k_n} \cap A_{k_n}^c} p_{\bar{Y}^{k_n}}(\mathbf{y}) + \sum_{\mathbf{y} \in \mathcal{P}_{k_n} \cap A_{k_n}} p_{\bar{Y}^{k_n}}(\mathbf{y}) \quad \text{(C.170b)}$$

$$\leq \frac{e^{-k_n\gamma}}{|\mathcal{P}_{k_n}|} \sum_{\mathbf{y} \in \mathcal{P}_{k_n} \cap A_{k_n}^c} 1 + \sum_{\mathbf{y} \in \mathcal{P}_{k_n} \cap A_{k_n}} p_{\bar{Y}^{k_n}}(\mathbf{y}) \quad \text{(C.170c)}$$

$$< e^{-k_n\gamma} + \epsilon \quad \text{(C.170d)}$$

showing it is impossible for $\overline{\mathbf{Y}} \in \mathcal{T}_0$. ∎

## APPENDIX D
## TYPE SET SIZE ENTROPY

*Theorem D.1:* Let $(p_1, p_2, \ldots)$ be a sequence of types defined over the finite alphabet $\mathcal{X}$ where $p_n \in \mathcal{P}_n$. Assume this sequence satisfies the following:

1) $p_n \to p$
2) $p_n \prec\prec p, \quad \forall n$

Then,

$$\lim_{n \to \infty} \frac{1}{n} \log |T(p_n)| = H(p). \quad \text{(D.171)}$$

*Proof:* We first show,

$$\liminf_{n \to \infty} \frac{1}{n} \log |T(p_n)| \geq H(p). \quad \text{(D.172)}$$

A sharpening of Stirling's approximation states that for $\frac{1}{12n+1} < \lambda_n < \frac{1}{12}$,

$$n! = \sqrt{2\pi} n^{n+\frac{1}{2}} e^{-n} e^{\lambda_n}.$$

Let the empirical distribution, $p_n$ be specified by $(n_1, \ldots, n_{K_n})$. If we enumerate the outcomes as $(a_1, \ldots, a_{K_n})$ we have that,

$$p_n(a_i) = \frac{n_i}{n}.$$

By definition $\sum_{i=1}^{K_n} n_i = n$, and from the above condition of absolute continuity we have that $K_n \leq s(p)$ for all $n$, where $s(p)$ is the support of the final distribution.

$$\log |T(p_n)| = \log \binom{n!}{n_1!, n_2!, \ldots, n_{K_n}!}$$

$$= \log \frac{\sqrt{2\pi} n^{n+\frac{1}{2}} e^{-n} e^{\lambda_n}}{\prod_{i=1}^{K_n} \left(\sqrt{2\pi} n_i^{n_i+\frac{1}{2}} e^{-n_i} e^{\lambda_{n_i}}\right)}$$

$$= n \log n - \sum_{i=1}^{K_n} n_i \log n_i + \log \sqrt{2\pi n} e^{\lambda_n}$$

$$- \sum_{i=1}^{K_n} \log \left(\sqrt{2\pi n_i} e^{\lambda_{n_i}}\right)$$

$$\geq n H(p_n) - K_n \log \left(\sqrt{2\pi n} e^{\frac{1}{12}}\right)$$

This implies that,

$$\frac{1}{n} \log |T(p_n)| \geq H(p_n) - \frac{s(p)}{n} \log \left(\sqrt{2\pi n} e^{\frac{1}{12}}\right).$$

Taking the $\liminf$ of each side,

$$\liminf_{n \to \infty} \frac{1}{n} \log |T(p_n)| \geq \liminf_{n \to \infty} H(p_n) = H(p). \quad \text{(D.173)}$$

Now we have from the type class upper-bound[14] that,

$$\limsup_{n \to \infty} \frac{1}{n} \log |T(p_n)| \leq \limsup_{n \to \infty} H(p_n). \quad \text{(D.174)}$$

Combing with (D.173) gives the desired result. ∎

## ACKNOWLEDGMENT

The support of the Center for Integrated Transmission and Exploitation (CITE) and the Information Directorate of the Air Force Research Laboratory, Rome, NY is gratefully acknowledged. The authors also wish to thank the anonymous reviewers for their helpful and constructive comments.

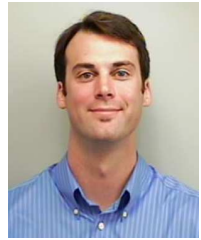

**Jeremiah Harmsen** is currently an engineer at Google Inc. in Mountain View, CA.

Jeremiah received a B.S. degree in electrical engineering and computer engineering (2001), a M.S. degree in electrical engineering (2003), a M.S. degree in mathematics (2005) and a Ph.D. in electrical engineering (2005) from Rensselaer Polytechnic Institute, Troy, NY.

Jeremiah received the Rensselaer Wynant James Prize in Electrical Engineering, Rensselaer Founders Award and Rensselaer Electrical Computer and Systems Departmental Service Award. He is a member of Tau Beta Pi and Eta Kappa Nu.

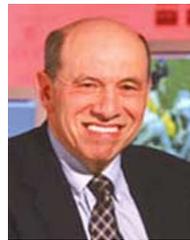

**William Pearlman** is Professor of Electrical, Computer and Systems Engineering and Director of the Center for Image Processing Research, Rensselaer Polytechnic Institute, Troy, NY.

Previously, he held industrial positions at Lockheed Missiles and Space Company and GTE-Sylvania before joining the Electrical and Computer Engineering Department, University of Wisconsin, Madison, WI, in 1974, and then moved to Rensselaer in 1979. He has spent sabbaticals at the Technion in Israel, Haifa, and Delft University of Technology, Delft, The Netherlands. He received an National Research Council associateship to spend six months in 2000 at the Air Force Research Laboratory, Rome, NY. In the summer of 2001, he was a Visiting Professor at the University of Hannover, Germany, where he was awarded a prize for outstanding work in the field of picture coding. His research interests are in data compression of images, video, and audio, digital signal processing, information theory, and digital communications theory.

Dr. Pearlman is a Fellow of SPIE. He has been a past Associate Editor for Coding for the IEEE TRANSACTIONS ONIMAGE PROCESSING and served on many IEEE and SPIE conference committees, assuming the chairmanship of SPIEs Visual Communications and Image Processing in 1989 (VCIP89). He was a keynote speaker at VCIP2001 and at the Picture Coding Symposium 2001 in Seoul, Korea. He has received the IEEE Circuits and Systems Society 1998 Video Transactions Best Paper Award and the IEEE Signal Processing Society 1998 Best Paper Award in the Area of Multidimensional Signal and Image Processing.